\documentclass[preprint]{aastex62}
\submitjournal{ApJ}
\shortauthors{Mace et al.}
\shorttitle{Wolf 1130}

\usepackage{graphicx, rotating, url}

\def\kms{\ifmmode{\rm km\thinspace s^{-1}}\else km\thinspace s$^{-1}$\fi}
\def\mps{\ifmmode{\rm m\thinspace s^{-1}}\else m\thinspace s$^{-1}$\fi}

\begin{document}

\title{Wolf 1130: A Nearby Triple System Containing a Cool, Ultramassive White Dwarf}

\author{Gregory N. Mace}
\affiliation{McDonald Observatory and Department of Astronomy, University of Texas at Austin, 2515 Speedway, Stop C1400, Austin, TX 78712-1205, USA}
\email{gmace@astro.as.utexas.edu}

\author{Andrew W. Mann}
\affiliation{McDonald Observatory and Department of Astronomy, University of Texas at Austin, 2515 Speedway, Stop C1400, Austin, TX 78712-1205, USA}

\author{Brian A. Skiff}
\affiliation{Lowell Observatory, 1400 West Mars Hill Road, Flagstaff, AZ 86001, USA}

\author{Christopher Sneden}
\affiliation{McDonald Observatory and Department of Astronomy, University of Texas at Austin, 2515 Speedway, Stop C1400, Austin, TX 78712-1205, USA}

\author{J.~Davy Kirkpatrick}
\affiliation{IPAC, Mail Code 100-22, Caltech, 1200 E. California Blvd. Pasadena, CA 91125}

\author{Adam C. Schneider}
\affiliation{School of Earth and Space Exploration, Arizona State University, Tempe, AZ, 85282}

\author{Benjamin Kidder}
\affiliation{McDonald Observatory and Department of Astronomy, University of Texas at Austin, 2515 Speedway, Stop C1400, Austin, TX 78712-1205, USA}

\author{Natalie M. Gosnell}
\affiliation{Department of Physics, Colorado College, 14 E. Cache La Poudre Street, Colorado Springs, CO 80903, USA}

\author{Hwihyun Kim}
\affiliation{Gemini Observatory, Casilla 603, La Serena, Chile}

\author{Brian W. Mulligan}
\affiliation{McDonald Observatory and Department of Astronomy, University of Texas at Austin, 2515 Speedway, Stop C1400, Austin, TX 78712-1205, USA}

\author{L. Prato}
\affiliation{Lowell Observatory, 1400 West Mars Hill Road, Flagstaff, AZ 86001, USA}

\author{Daniel Jaffe}
\affiliation{McDonald Observatory and Department of Astronomy, University of Texas at Austin, 2515 Speedway, Stop C1400, Austin, TX 78712-1205, USA}

\begin{abstract}

Following the discovery of the T8 subdwarf WISE~J200520.38$+$542433.9 (Wolf~1130C), with common proper motion to a binary (Wolf~1130AB) consisting of an M subdwarf and a white dwarf, we set out to learn more about the old binary in the system. 
We find that the A and B components of Wolf~1130 are tidally locked, which is revealed by the coherence of more than a year of V band photometry phase folded to the derived orbital period of 0.4967 days. 
Forty new high-resolution, near-infrared spectra obtained with the Immersion Grating Infrared Spectrometer (IGRINS) provide radial velocities and a projected rotational velocity ($v$~sin~$i$) of 14.7 $\pm$ 0.7 \kms\ for the M subdwarf. 
In tandem with a {\it Gaia} parallax-derived radius and verified tidal-locking, we calculate an inclination of i$=$29$\pm$2 degrees. 
From the single-lined orbital solution and the inclination we derive an absolute mass for the unseen primary ($1.24^{+0.19}_{-0.15}$ M$_{\odot}$).
Its non-detection between 0.2 and 2.5$\mu$m implies that it is an old ($>$3.7 Gyr) and cool (T$_{\rm eff}$ $<$7000K) ONe white dwarf. 
This is the first ultramassive white dwarf within 25~pc.
The evolution of Wolf 1130AB into a cataclysmic variable is inevitable, making it a potential Type~Ia supernova progenitor. The formation of a triple system with a primary mass $>$100 times the tertiary mass and the survival of the system through the common-envelope phase, where $\sim$80$\%$ of the system mass was lost, is remarkable. Our analysis of Wolf 1130 allows us to infer its formation and evolutionary history, which has unique implications for understanding low-mass star and brown dwarf formation around intermediate mass stars.\\

\end{abstract}


\section{Introduction}

Extreme mass-ratio systems with brown dwarf companions test our understanding of star formation and evolution \citep{bate2003,bate2009} because they are rare \citep{parker2013,derosa2014}.
While most intermediate-mass stars are binaries or multiples \citep{preibisch1999,garcia2001}, stars more massive than the Sun dominate star forming regions and quench nearby core fragmentation and accretion \citep{zinnecker2007}, which then impacts brown dwarf formation. 
The most likely formation path for massive brown dwarfs is core collapse, just like stars above the hydrogen burning limit \citep{bate2002}. 
Yet, brown dwarfs may also form through fragmentation of massive circumstellar disks \citep{bonnell1994,kratter2006,stamatellos2011}.
The characterization of brown dwarfs in multiple systems, with extreme mass ratios and unique orbital parameters, is required to identify the bounds of stellar formation models \citep{bate2009}.

The greatest limitation to discovering low-mass companions to high-mass stars is the luminosity contrast.
Faint brown dwarfs are easily hidden from observation since the mass-luminosity relation for main-sequence stars scales with luminosity approximately as the mass to the 4th power. Hence, a mass ratio $>$100 results in a luminosity ratio $>$10$^8$. 
T-type brown dwarfs in multiple systems are also rare, with only five examples out of $>$550 known T dwarfs \citep[DwarfArchives.org\footnote{\url{http://spider.ipac.caltech.edu/staff/davy/ARCHIVE/index.shtml}};][]{mace2013a,mace2014,deacon2017}, while theoretical calculations hint at an overall multiplicity rate of $\sim$10$\%$ \citep{bate2009}.
Imaging \citep{derosa2014} and spectroscopic studies \citep{gullikson2016b} of intermediate-mass stars have recently added to the sample of high-contrast binaries, but most observable companions are not substellar.
The detectability of low-mass companions is improved once the more massive star becomes a white dwarf, the system mass is reduced \citep{burleigh2011}, and the orbital separation increases while the flux contrast decreases.

Wolf~1130 (Gl~781, LHS~482, HIP~98906) is a nearby \citep[16.7$\pm$0.2~pc,][]{gaia2016, gaia2016DR1} M subdwarf (Wolf~1130A) and white dwarf (Wolf~1130B) binary with a 0.4967 day orbital period \citep{gizis1998}. It is also classified as a flare star with the designation V1513 Cyg.
Wolf~1130C is a $\sim$800K subdwarf brown dwarf with a common proper motion to Wolf~1130AB and a projected separation of $\sim$3150~AU \citep{mace2013b}.
The now abundant sample of late-type T dwarfs have similar surface temperatures to Wolf~1130C but are not direct counterparts (Logsdon et al. submitted).
\citet{derosa2014} seem to have found a young version to Wolf~1130C as a mid-type L dwarf with a mass of $\sim$0.050~M$_{\odot}$ around an intermediate mass star.
Yet, Wolf~1130C is unique to the sample of benchmark T dwarfs because it is the oldest of the five higher order multiple systems \citep{deacon2017} and is distinctly on the edge of model parameter space with the lowest metallicity, a small radius, high mass and large surface gravity \citep{mace2013b}.
This triple system is useful for understanding the evolution of intermediate-mass stars, close binaries, red dwarfs, and brown dwarfs.

White dwarf and M dwarf eclipsing binaries in the literature have similar 0.5-1.5 day orbital periods to Wolf~1130AB and mass ratios near unity \citep{maxted2004, muirhead2013}. 
V471~Tau is similar to Wolf~1130, both systems containing a low-mass star orbiting a white dwarf in a $\sim$0.5~day period with a companion near the substellar boundary \citep{Vaccaro2015}, but Wolf~1130 is old and V471~Tau is a member of the Hyades \citep[$\sim$800~Myr,][]{Brandt2015}.
The rapid rotation of the M dwarf in V471~Tau drives significant starspots \citep{Kundra2011}, which likely exist on Wolf~1130A at much smaller scales.
Wolf~1130 is not eclipsing, but by estimating the M subdwarf radius and measuring $v$~sin~$i$ we can determine the inclination and derive the absolute mass for the unseen white dwarf.
We find that Wolf~1130AB is an evolved, pre-cataclysmic version of the intermediate-mass star and M dwarf binaries that \citet{gullikson2016b} characterized.
Additionally, this is the nearest ultramassive white dwarf \citep{Cummings2016}, and a potential Type~Ia supernova progenitor.

\section{Observations} 

\subsection{Optical Photometry}
We obtained {\it V}-band CCD photometry of Wolf~1130AB using the Lowell 31-inch (0.7-m effective aperture) telescope in robotic mode on 61 nights between UT dates 2014 July 19 and 2015 November 03.  
Usually several visits were made each night with the field above 2.5 airmasses, but on several occasions continuous observations were obtained in hopes of resolving the period alias near one day.
A total of 1170 observations were obtained. The data were reduced via aperture photometry using four comparison stars in the 15'x15' field and the magnitude zero-point was adjusted approximately to standard {\it V}.

\begin{figure}
\epsscale{0.9}
\figurenum{1}
\centering
\includegraphics[width=5in]{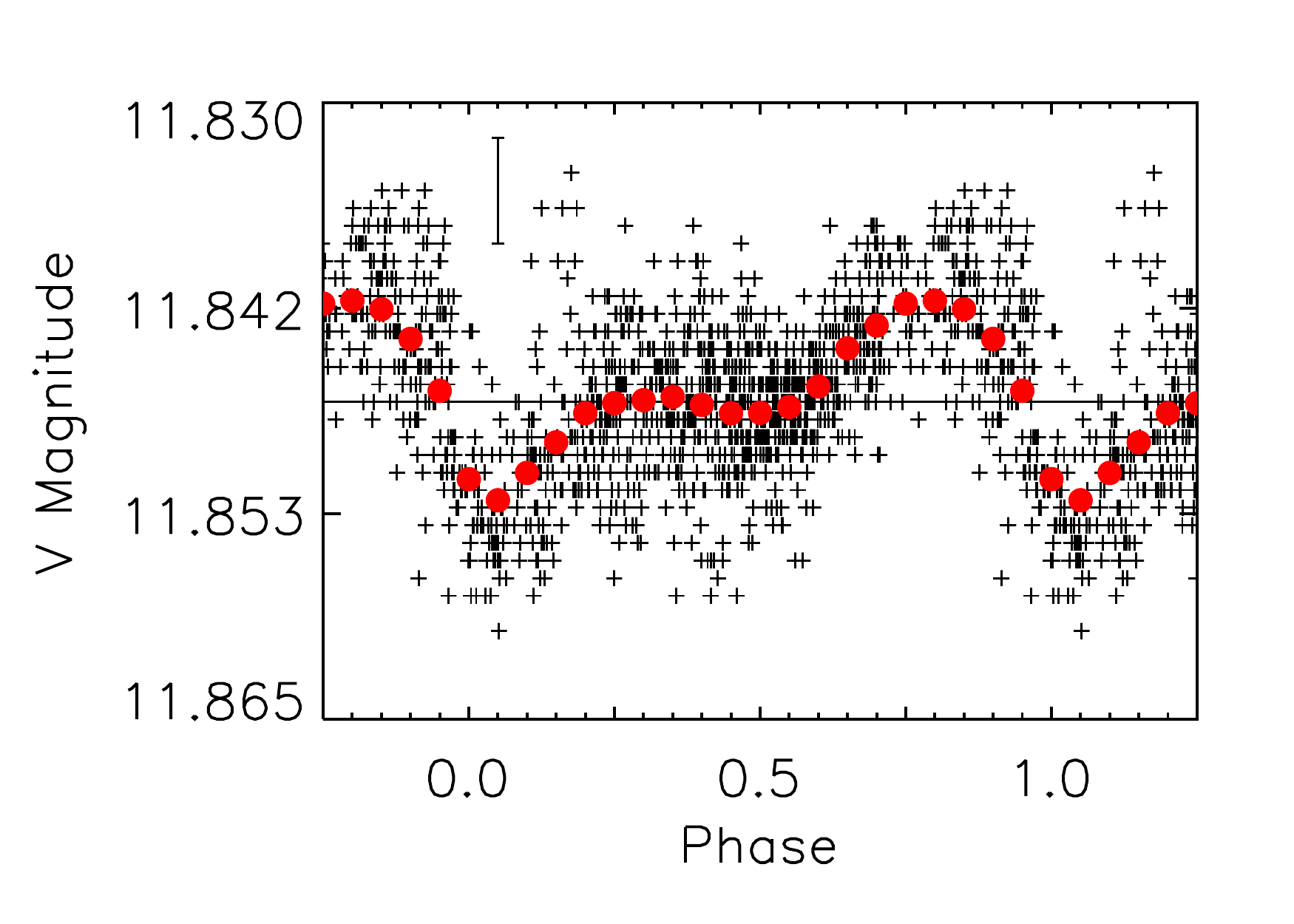}
\caption{{\it V}-band photometric measurements of Wolf~1130AB phase folded to the orbital period of 0.4967~days. The average relative uncertainty, 0.003 magnitudes, is shown in the upper corner. Binned photometry is shown for every 0.05 phase with bin widths of 0.1 phase (red circles). The horizontal line marks the average magnitude. We see no outbursts or eclipses in our {\it V}-band observations across 472 days of observation. 
\label{phot}}
\end{figure}

A periodogram of the photometry constructed using the NASA Exoplanet Archive Periodogram Tool\footnote{\url{https://exoplanetarchive.ipac.caltech.edu}} and the \citet{plavchan2008} algorithms reveals two major photometric periods. The primary peak in the periodogram is at 0.9885 $\pm$ 0.0003 days. However, this $\sim$1 day period clusters most photometry at phases between 0.4-0.9.\footnote{In this work phase$=$0 at inferior conjunction (M subdwarf between the observer and the white dwarf).} The second peak in the periodogram is similar in power to the primary peak and provides a period of 0.4966 $\pm$ 0.0001 days, which is consistent with the orbital period. 
Figure~\ref{phot} shows the Wolf~1130AB {\it V}-band photometry, listed in Table~\ref{tbl-1}, phase-folded to the orbital period of the system. The average relative uncertainty in the photometry is 0.003 magnitudes and the binned photometry has errors on the order of the symbol size (0.0003 mag.). 
The coherence of the photometry when phase-folded to the orbital period and corrected for orbital effects is consistent with Wolf~1130AB being tidally locked, which we discuss more in Section 4.1.

More than 750 observations of Wolf~1130 are included in the ASAS-SN database \citep{Shappee2014,Kochanek2017}\footnote{\url{https://asas-sn.osu.edu/}}. The real-time aperture extraction operates at fixed, user-provided coordinates and the brightness of Wolf~1130 decreases with time ($\sim$0.4 magnitudes over $\sim$3.5 years). However, this decrease could be due to variability in the star or the changing position ($\sim$1$\farcs$5 per year) of Wolf~1130 relative to the fixed aperture. Some of the ASAS-SN photometric measurements are significantly higher than the baseline, supporting the flare star designation for Wolf 1130, although we see no outbursts or eclipses in our {\it V}-band observations over 472 days. 
We do not consider the ASAS-SN photometry further in our analysis since the uncertainties are more than twice the amplitude of variation shown in Figure~\ref{phot}.

\subsection{Optical Spectroscopy}
\begin{figure}
\epsscale{0.9}
\figurenum{2}
\centering
\includegraphics[width=5in]{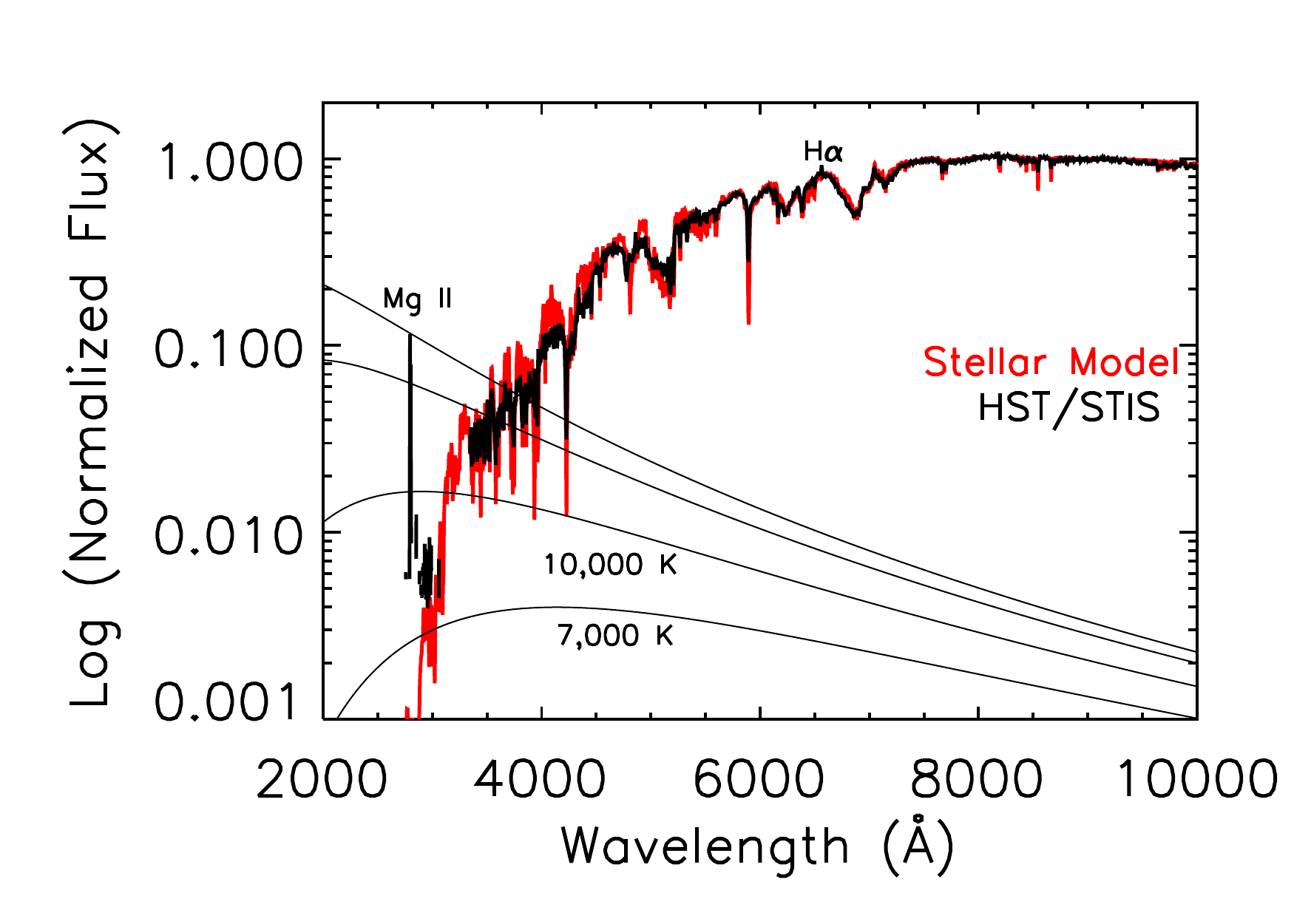}
\caption{The HST/STIS spectrum of Wolf~1130AB where signal-to-noise is $>$5. Mg II and H$\alpha$ emission are marked. The inclusion of the best fit BT-SETTL model \citep{allard2014} with T$_{\rm eff}$=3500~K, log~$g$=4.83, and [Fe/H]$=-$1.3 illustrates the absence of flux from the more massive companion. Scaled Planck functions for 7,000, 10,000, 15,000 and 20,000~K black bodies are included for comparison. As discussed in Section 4.2, we find the temperature of the massive companion to be $\leq$7000K when we assume a white dwarf radius of 0.005~R$_{\odot}$.
\label{optical_spec}}
\end{figure}

\begin{sidewaysfigure*}
\epsscale{1.1}
\figurenum{3}
\centering
\includegraphics[width=5in]{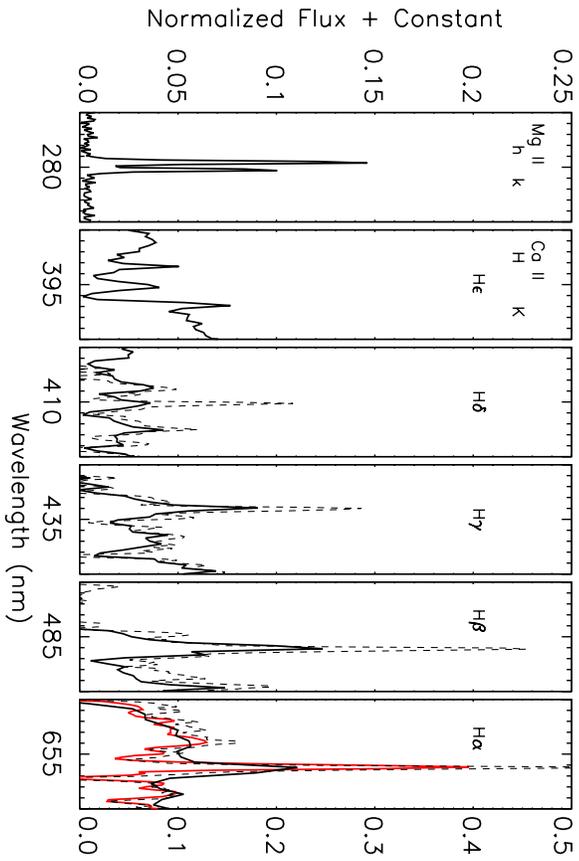}
\caption{Emission lines in the Wolf~1130AB spectra observed with HST/STIS (solid line), Palomar/DoubleSpec (dashed line), and MDM (red line). Each panel is 10~nm wide and the flux scale on the right is for the H$\alpha$ panel only. Although variable, \citet{gizis1998} find no correlation of the emission with orbital phase.
\label{optical_emission}}
\end{sidewaysfigure*}

Wolf~1130AB was observed on 2003 August 10 UT (JD 2452861.815015\footnote{JD and MJD in this work are standard, and not heliocentric corrected.}) with the Space Telescope Imaging Spectrograph (STIS) on the Hubble Space Telescope (HST) as part of the Next Generation Spectral Library program \citep{gregg2006} and is publicly available on The Mikulski Archive for Space Telescopes (MAST).
The STIS spectrum covers wavelengths between 1675 and 10196~\AA\ with a spectral resolution of $\sim$4~\AA. 
Another optical spectrum of Wolf~1130AB from the Mark III spectrograph at MDM Observatory was taken on 2011 May 08 UT (JD 2455689.995729). The spectrum was originally part of the catalog of low-mass stars characterized by \citet{gaidos2014} and covers between 6200 and 8300~\AA\ at a resolution of 5.4~\AA.
A third optical spectrum was acquired with DoubleSpec at Palomar Observatory on 2014 June 25 UT (JD 2456833.884028). The DoubleSpec blue channel covers 4100 to 7100~\AA\ at a resolution of 3.5~\AA\ and we do not use the red channel data in this work because of telluric contamination and poor flux normalization.
Figure~\ref{optical_spec} shows the HST/STIS spectrum, which has the highest signal-to-noise ratio and broadest spectral coverage out of the three optical spectra. 
The prominent emission features observed in each optical spectrum are highlighted in Figure~\ref{optical_emission}. \citet{gizis1998} finds no correlation of the H$\alpha$ emission with orbital phase.

\subsection{Infrared Spectroscopy}
A near-infrared spectrum of Wolf~1130AB was obtained using the SpeX spectrograph \citep{rayner2003} on the NASA Infrared Telescope Facility (IRTF) on 2013 May 15 UT. Employing the short cross-dispersed mode and the 0$\farcs$3$\times15\arcsec$ slit, we obtained simultaneous coverage from 0.7 to 2.5$\mu$m at $R\simeq2000$. The target was observed at two positions along the slit for sky background subtraction. Six pairs were taken in total. The spectrum was flat fielded, extracted, wavelength calibrated, and stacked using the \textit{SpeXTool} package \citep{cushing2004}, which provided a final S/N $>$150 per resolution element in the $H$-band. An A0V-type star was observed immediately after the target and was used for telluric correction using the \textit{xtellcor} package \citep{vacca2003}.

Wolf~1130AB was observed with IGRINS (Immersion Grating Infrared Spectrometer) 40 times between 2014 July 11 and 2015 August 05 UT at McDonald Observatory. A log of the observations is provided in Table~\ref{tbl-2}. IGRINS is unique in its ability to observe the entire H and K bands (1.45-2.5$\mu$m) in a single exposure at R$\approx$45,000 \citep{yuk2010, park2014, mace2016b}. Each resolution element is $\sim$7~\kms\ and there are more than 20,000 resolution elements in a single IGRINS spectrum. At the heart of IGRINS is a silicon immersion grating \citep{marsh2007,gully2012}, which amplifies the dispersion by the index of refraction \citep[n$\approx$3.4 for silicon at 130K, ][]{frey2006}. The IGRINS data reduction pipeline \citep{jlee2016}\footnote{\url{https://github.com/igrins/plp/tree/v2.1-alpha.3}} employs flat lamps from the IGRINS calibration unit to derive pixel variance, night-sky OH emission and telluric absorption lines for wavelength calibration, and the optimal extraction methods of \citet{horne1986} to produce 1-D spectra. The final pipeline output consists of 44 spectral orders (23 in H and 21 in K) with $\sim$10$\%$ spectral overlap between orders. Telluric absorption relies on division by an A0V star observed at a similar airmass to each observation and multiplied by the \citet{kurucz1979,kurucz11}\footnote{\url{http://kurucz.harvard.edu/stars/vega/}} Vega model. 

\begin{sidewaysfigure*}
\epsscale{1}
\figurenum{4}
\centering
\includegraphics[width=6in,angle=90]{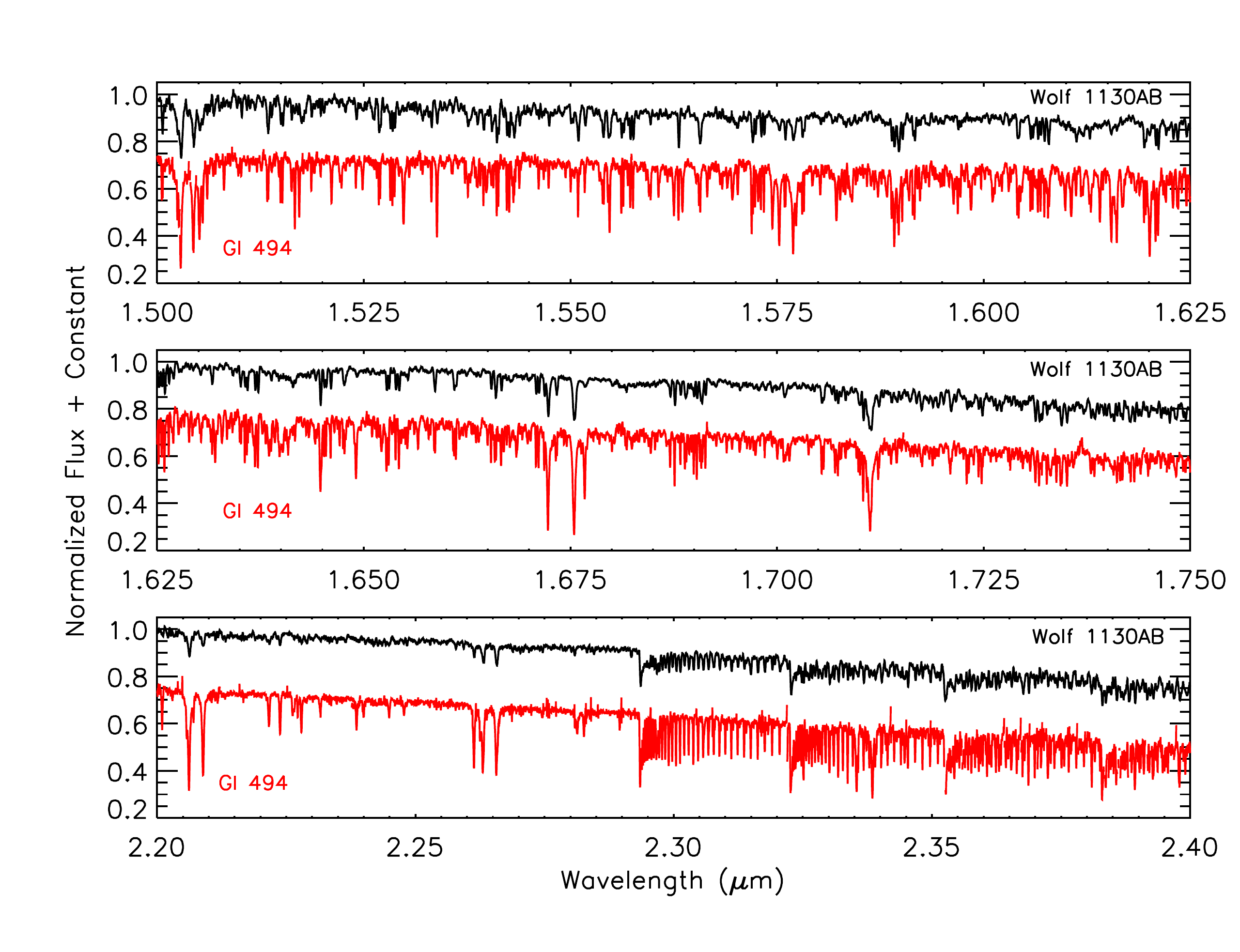}
\caption{IGRINS spectra of Wolf~1130AB and Gl~494 shifted to a common wavelength. Both have spectral types of $\sim$M2 and comparable rotational velocities, but Gl~494 has super-solar metallicity, while Wolf~1130A is a subdwarf.
\label{IGRINS_comp}}
\end{sidewaysfigure*}

\begin{figure}
\epsscale{1.0}
\figurenum{5}
\centering
\includegraphics[width=5in]{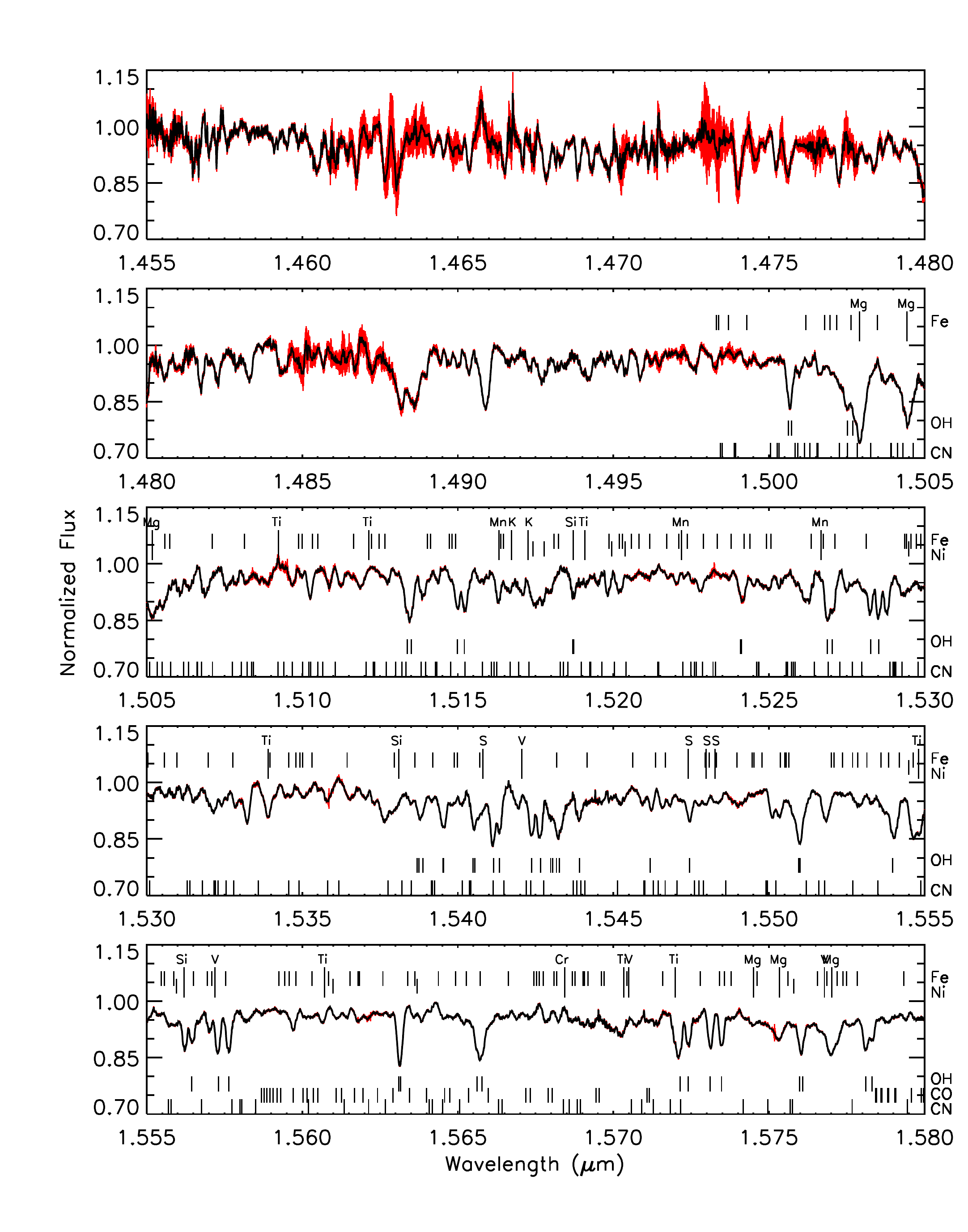}
\caption{Combined IGRINS spectrum of Wolf~1130AB (black line) with 1-$\sigma$ uncertainties (red region). By velocity shifting and median combining 40 epochs, a signal-to-noise $\sim$650 spectrum is produced which spans the entire H and K atmospheric windows. Line identifications from the Arcturus atlas \citep{hinkle1995} and HITRANonline \citep{HITRAN2012} are labeled on the plot. Atomic lines that are too abundant to label above the spectrum (Fe and Ni) are labeled on the right-hand margin with marks above the spectrum. Molecular lines (OH, CO, and CN) are also labeled in the right-hand margin, but with identification marks below the spectrum.
\label{IGRINS_spec1}}
\end{figure}

All 40 IGRINS spectra have been combined into a single spectrum by shifting to a common wavelength and then re-binning the flux in 0.00001$\mu$m bins while removing 3$\sigma$ outliers. The uncertainty in the combined spectrum is the standard deviation of the mean of all flux measurements in a bin. The resultant spectrum has an average uncertainty of $\sim$0.15$\%$ (signal-to-noise$\sim$650). In Figure~\ref{IGRINS_comp} we compare Wolf~1130AB to the field star Gl~494 (T$_{\rm eff}$=3570 K, $v$~sin~$i$=10 \kms, [Fe/H]$\sim+$0.2 dex, log~$g$=4.5 dex; \citet{jenkins2009, lepine2013, neves2013}), which was observed with IGRINS on 2015 April 02 UT. 
Wolf~1130A and Gl~494 are essentially the same temperature and have similar $v$~sin~$i$, but differences in metallicity and surface gravity produce contrasting line depths. 
The entire combined IGRINS spectrum of Wolf~1130AB is presented in Figure~\ref{IGRINS_spec1} with lines identified using the Arcturus atlas of \citet{hinkle1995} and molecular line lists from HITRANonline \citep{HITRAN2012}. 
Most unmarked lines are H$_2$O and too numerous to label. 
We do not measure any variation in the line profiles or depths between epochs.
The remaining uncertainties in the combined spectrum are primarily a result of variance in the telluric absorption between epochs of observation. 

\clearpage

\clearpage

\begin{figure}
\epsscale{1.0}
\figurenum{5}
\centering
\includegraphics[width=6in]{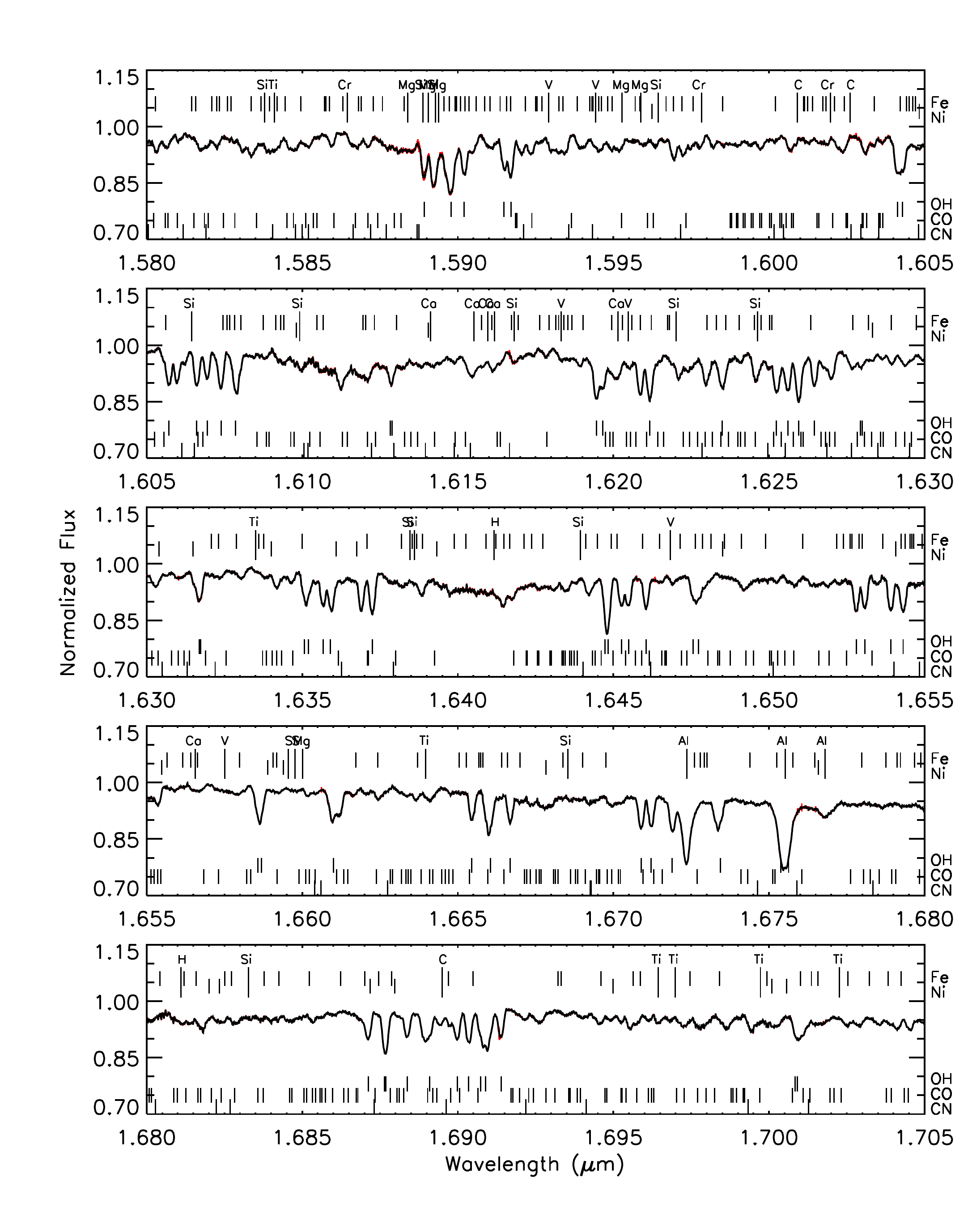}
\caption{Combined IGRINS spectrum of Wolf~1130AB - Continued.
\label{IGRINS_spec2}}
\end{figure}

\clearpage

\begin{figure}
\epsscale{1.0}
\figurenum{5}
\centering
\includegraphics[width=6in]{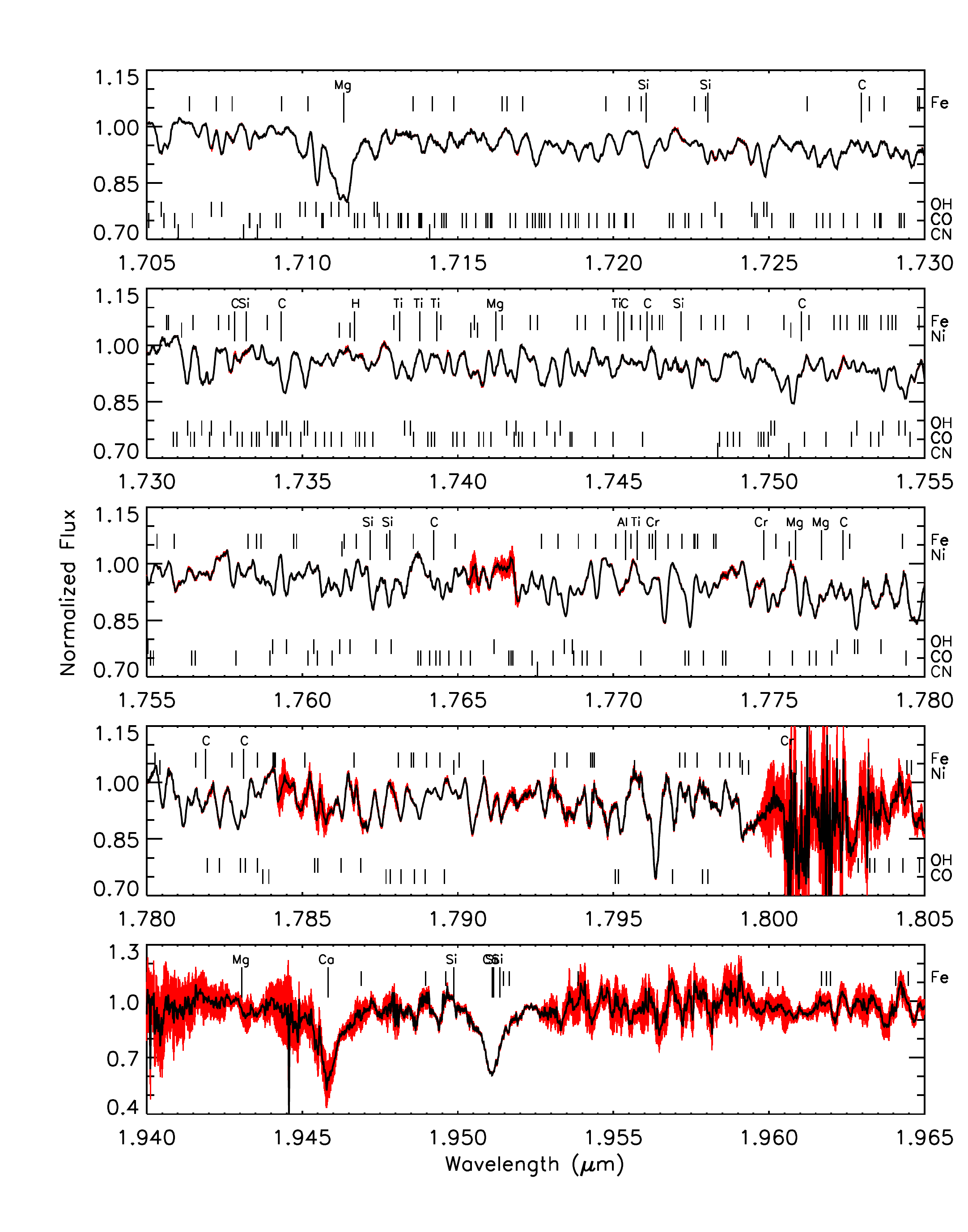}
\caption{Combined IGRINS spectrum of Wolf~1130AB - Continued.
\label{IGRINS_spec3}}
\end{figure}

\clearpage

\begin{figure}
\epsscale{1.0}
\figurenum{5}
\centering
\includegraphics[width=6in]{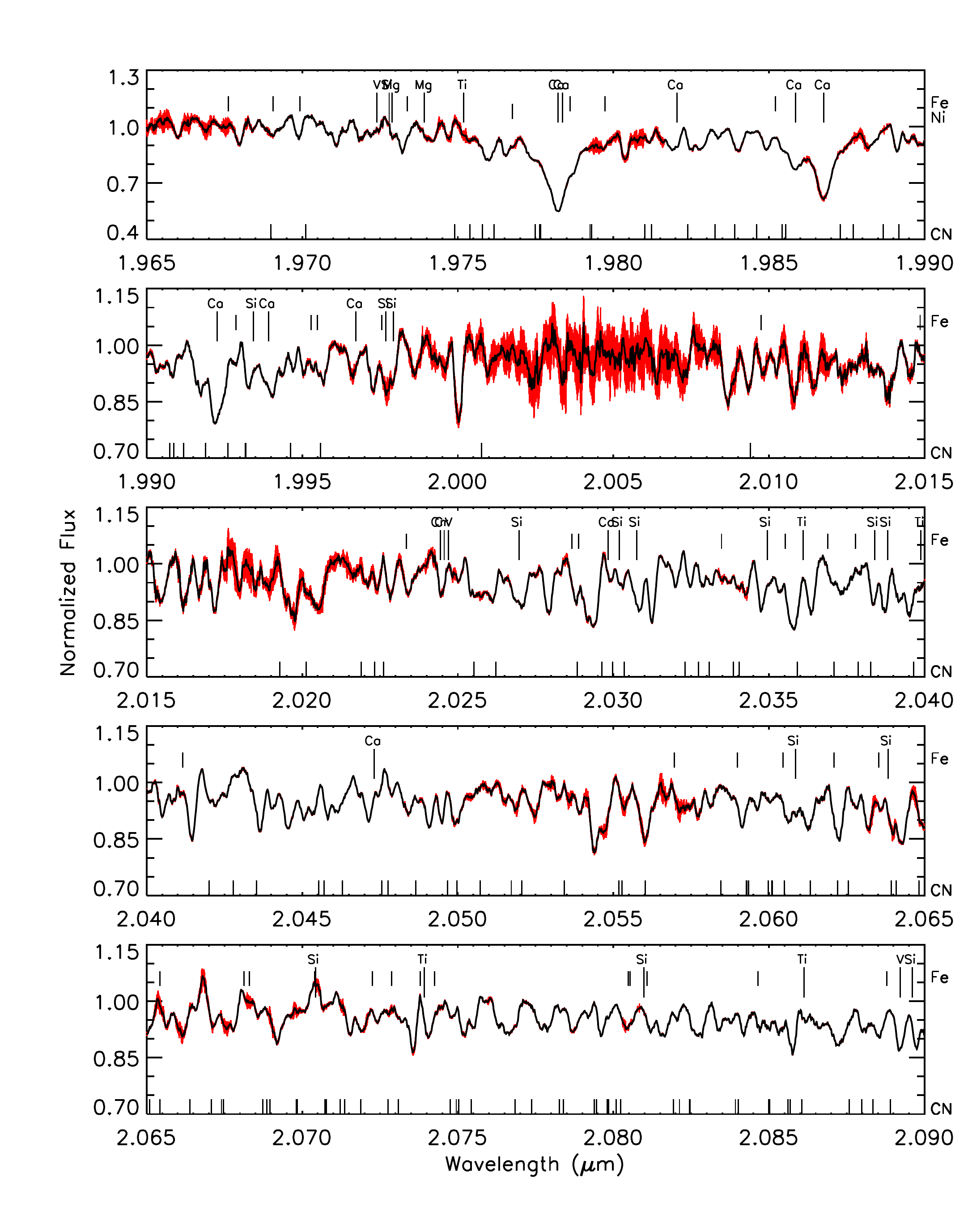}
\caption{Combined IGRINS spectrum of Wolf~1130AB - Continued.
\label{IGRINS_spec4}}
\end{figure}

\clearpage

\begin{figure}
\epsscale{1.0}
\figurenum{5}
\centering
\includegraphics[width=6in]{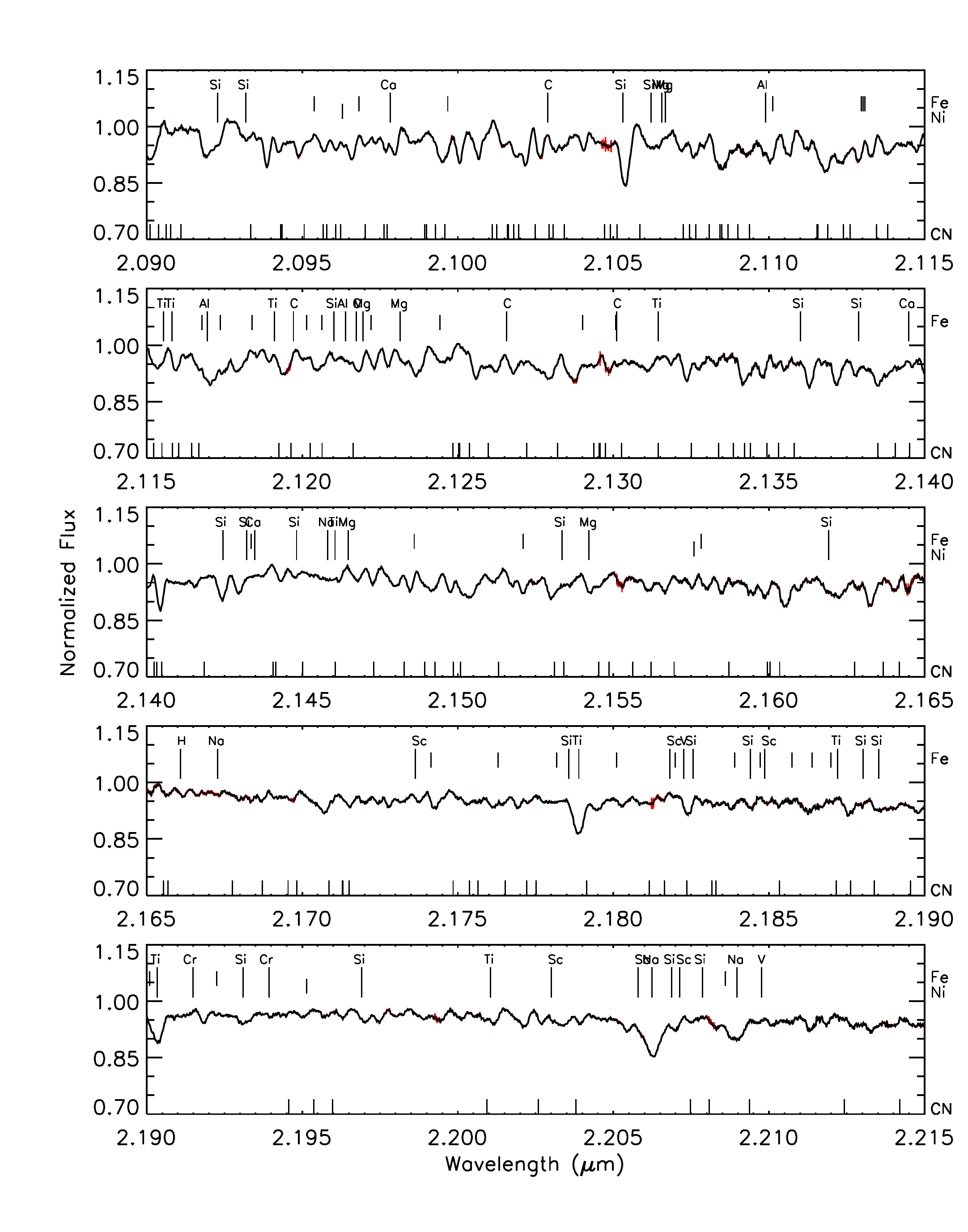}
\caption{Combined IGRINS spectrum of Wolf~1130AB - Continued.
\label{IGRINS_spec5}}
\end{figure}

\clearpage

\begin{figure}
\epsscale{1.0}
\figurenum{5}
\centering
\includegraphics[width=6in]{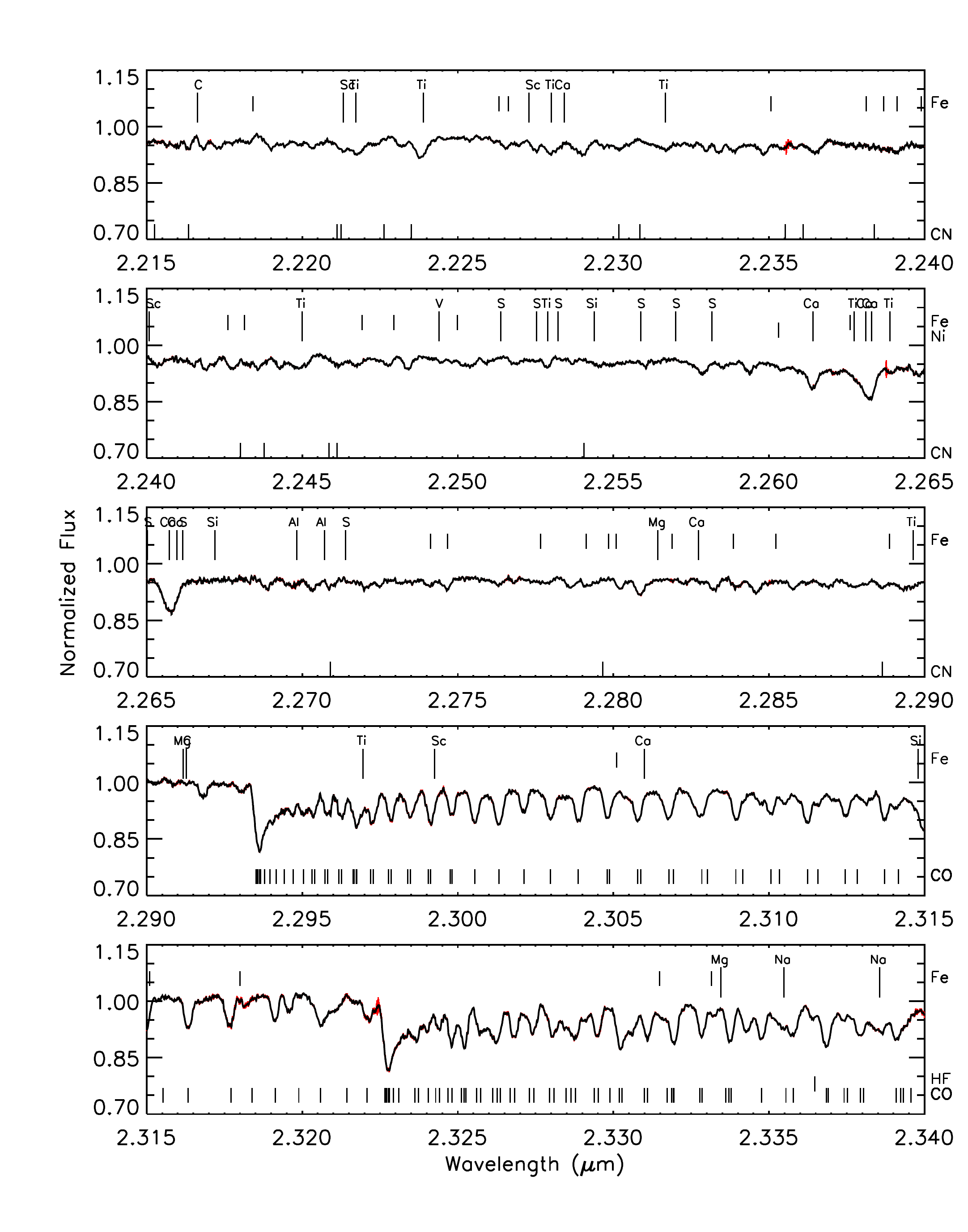}
\caption{Combined IGRINS spectrum of Wolf~1130AB - Continued.
\label{IGRINS_spec6}}
\end{figure}

\clearpage

\begin{figure}
\epsscale{1.0}
\figurenum{5}
\centering
\includegraphics[width=6in]{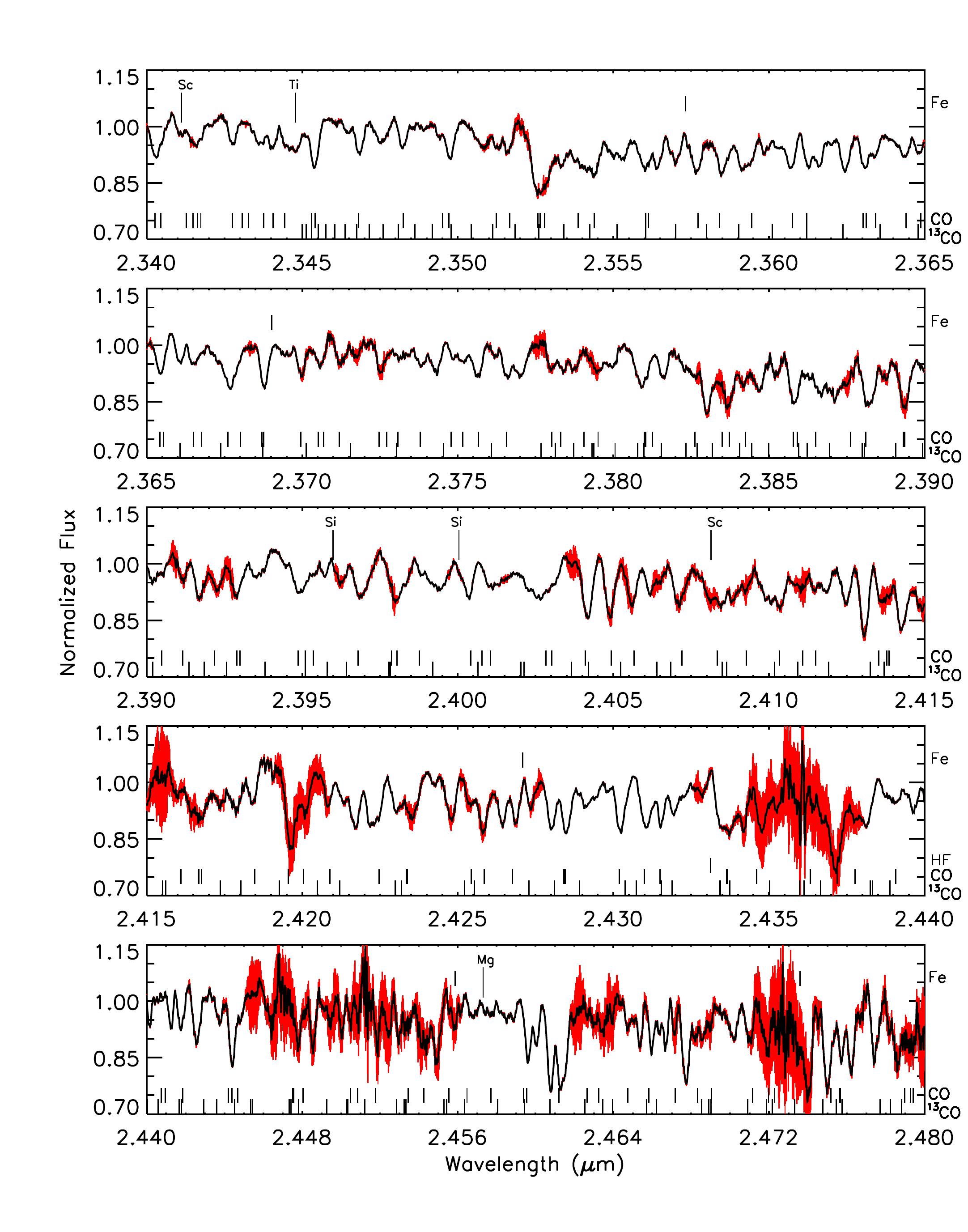}
\caption{Combined IGRINS spectrum of Wolf~1130AB - Continued.
\label{IGRINS_spec7}}
\end{figure}

\clearpage

\subsection{Parallax Measurements}
There are a number of parallax measurements in the literature for Wolf~1130 that place it between 14.9 and 17.5~pc from the Sun \citep{harrington1980,vanaltena1995,vanleeuwen2007}.
{\it Gaia} Data Release 1 \citep[DR1, ][]{gaia2016, gaia2016DR1,lindgren2016} adds a new parallax to the sample, $\pi$ = 59.91$\pm$0.55 mas, which is consistent with, and more precise than, previous measurements. 
The distance to Wolf~1130 from the {\it Gaia} DR1 parallax is 16.7$\pm$0.2~pc, which we adopt for all our analysis.

\section{Orbital Properties}

\subsection{Radial Velocity Determination}
\citet{gizis1998} derived 27 radial velocities for Wolf~1130A by cross-correlating 0.3~\AA\ resolution visible-light spectra (4700 to 9600~\AA). 
Most of the observations from \citet{gizis1998} were obtained on two adjacent nights in 1996 August and have typical uncertainties of $\pm$2 \kms. 
The combination of those observations with 40 IGRINS epochs produces a baseline of almost 20 years. 

IGRINS radial velocities were derived using the method summarized in \citet{mace2016a}.
For each of the 40 epochs of observation, the 44 orders of IGRINS spectra were cross-correlated against the other epochs and 185 other M dwarfs observed with IGRINS. 
The relative velocities were converted to absolute velocities by determining the zero-point offset relative to radial velocities in the literature \citep{nidever2002, maldonado2010, shkolnik2012,chubak2012, naud2014} for 103 of the other M dwarfs. 
This method provides radial velocities that are precise to $\sigma$ $=$ 160~\mps, with the uncertainty primarily set by the zero-point uncertainty in the literature for the M dwarf sample. 
Table~\ref{tbl-2} lists the radial velocities for each IGRINS observation of Wolf~1130AB.

\subsection{Orbital Solution}
The orbital parameters in Table~\ref{tbl-3} were determined by fitting the visible-light and infrared radial velocities separately and also as a combined set.
We used the Systemic Console 2 software package \citep{meschiari2009,meschiari2010} to find the orbital solution and its uncertainties with the constant assumption that Wolf~1130A has a mass of 0.3~M$_{\odot}$ (this assumption is motivated in Section 4.1). 
The best fit we find to the 27 radial velocities in \citet{gizis1998} is most similar to solution C in Table 3 of that paper. 
The most notable aspect of the fit to the visible-light radial velocities is the non-zero eccentricity, which is surprising given the systems old age and short orbital period.

A similar fit to the 40 infrared radial velocities alone produces a couple of notable differences relative to the visible-light fit. One difference is that the reduced chi-squared value for the fit is much larger because of the order of magnitude smaller uncertainties on the IGRINS residual radial velocities. 
Next, the eccentricity is consistent with zero in the infrared-only fit. The implication of an eccentricity of zero in the infrared and non-zero in the visible-light data is consistent with the {\it V}-band variations we observe, which are discussed in the next section on radial velocity residuals. Changes in the other orbital parameters are within the uncertainties and consistent with the larger reduced chi-square of the infrared-only fit.

\begin{figure}
\epsscale{0.9}
\figurenum{6}
\centering
\includegraphics[width=5in]{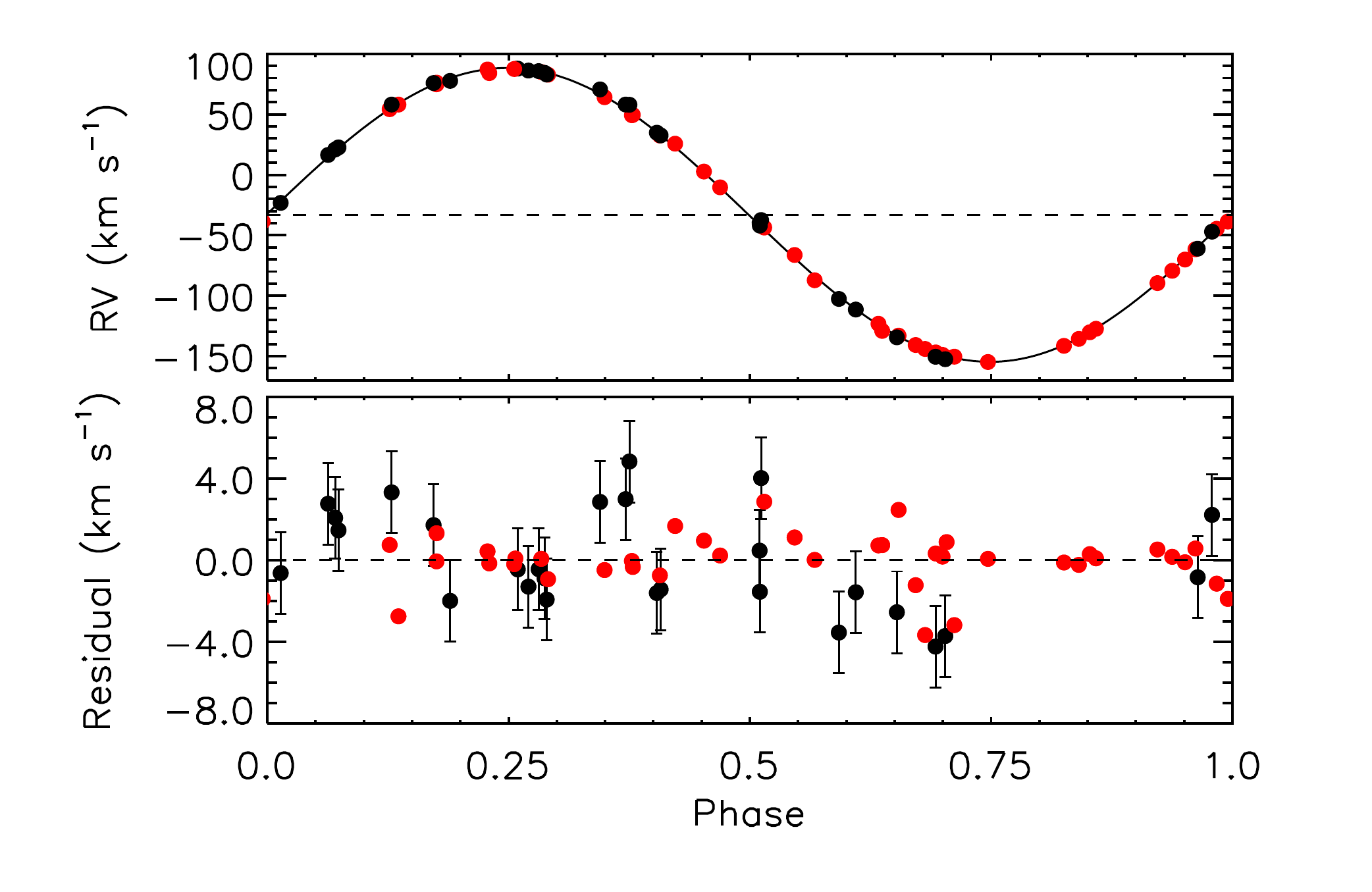}
\caption{Top: The phase folded radial velocities for the M subdwarf, Wolf 1130A, and the derived radial velocity curve for the combined visible-light and infrared observations. Our preferred orbital solution combines 67 radial velocities, fixed values for eccentricity (0) and inclination (29$^{\circ}$), and is provided in Table~\ref{tbl-3}. Visible-light radial velocities are from \citet[][black points]{gizis1998} and infrared radial velocities are from this work (red points). Bottom: The combined velocity residuals for the visible-light (black) and infrared (red) observations with eccentricity fixed at zero.  The standard deviations of the velocity residuals are $\sigma=$ 2.5 \kms\ for the visible-light data and $\sigma=$ 1.3 \kms\ in the infrared.
\label{orbit}}
\end{figure}

Combining the visible-light and infrared radial velocities improves the precision on the orbital period by an order of magnitude. The larger number of infrared radial velocities, in combination with their smaller uncertainties, causes the combined fit to be most consistent with the infrared-only solution. The combined visible light and infrared orbital solution for Wolf~1130AB, where only the mass of Wolf 1130A is fixed and all other parameters are determined, is shown in Figure~\ref{orbit}.

\subsection{Radial Velocity Residuals}
\begin{figure}
\epsscale{0.9}
\figurenum{7}
\centering
\includegraphics[width=5in]{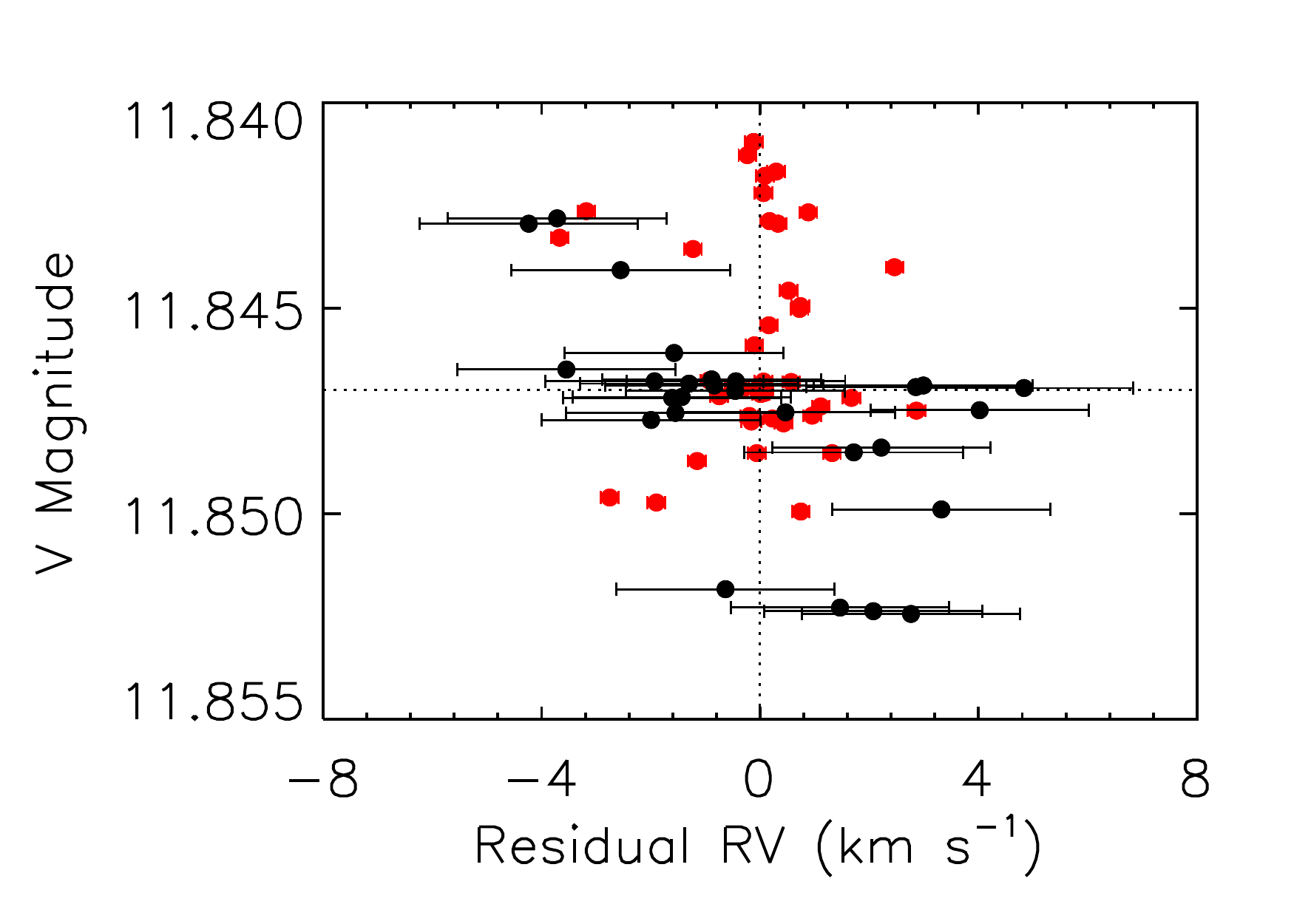}
\caption{Binned photometry plotted against the combined-orbital-solution radial velocity residuals for the same phases. Visible-light (black) and infrared (red) observations both show variation, but with differing amplitudes. The trend for the visible-light radial velocity residuals show a blue shift when the M subdwarf is brighter and a red shift when the M dwarf is fainter.
\label{residual_mag}}
\end{figure}

As an active star, measurements of Wolf~1130A are expected to behave like a young star and show radial velocity jitter on the order of a few 100~\mps\ at infrared wavelengths, but as high as 2~\kms\ in visible-light spectra \citep{mahmud2011}.
Velocity residuals for the combined orbital solution of Wolf~1130, with e=0, are shown in Figure~\ref{orbit} and are larger in the visible-light data ($\sigma=$ 2.5 \kms) than the infrared ($\sigma=$ 1.3 \kms).
Differences in the amplitude of variation implies that temperature variations are present on the stellar surface \citep{prato2008} at a 2:1 ratio that is similar to T Tauri stars \citep{crockett2012}. 
Figure~\ref{residual_mag} compares the binned photometry to the velocity residual at the same phases and reveals a trend.
Overall, the visible-light radial velocities are blue shifted when the M subdwarf is brighter and red shifted when it is fainter. 
This pattern holds true over the $\sim$20 year baseline between spectroscopic observations and for the {\it V}-band photometry presented here. 
Though smaller than the visible-light residuals, the IGRINS infrared radial velocity residuals are non-zero and show amplitude variations as a function of time which would require long-lived, phase-dependent sources of variation.
As we discuss in Section~4.1, surface temperature gradients, tidal elongation of the M dwarf, and Doppler beaming effects cause variability that are fixed with phase. 
These effects can also produce an offset photocenter, which explains part of the {\it V}-band variability and the non-zero eccentricity derived from the visible-light data. 

\section{Physical Parameters}

\citet{mace2013b} presented the discovery of Wolf~1130C and infrared magnitudes of the Wolf~1130AB system can be found in Table 1 of that paper. Newly determined parameters for Wolf~1130 are listed in Table~\ref{tbl-4} of this work.

\subsection{Wolf 1130A - M Subdwarf}

The radial velocity variablity of Wolf~1130A was first identified by \citet{Joy1947}. 
Standard practice is to define the brightest component of a system as the A component \citep{Hartkopf2004}, however, in a mass donor scenario like a cataclysmic variable mass is transferred from the secondary to the more massive primary \citep{Ritter1998}.
In the Wolf~1130 system the most massive component is the white dwarf, but it remains undetected and we call it Wolf~1130B. 
Wolf~1130A is the M subdwarf that will become the mass donor of the system and is the most luminous component.

\begin{itemize}

\item \textbf{Tidal Locking - }

There are three primary components of light curve modulation created by the influence of a close companion \citep{shporer2017}. The phase dependence of each of these are shown in Figure~\ref{phot_fits}. 
We determine the ellipsoidal variability (1.5$\pm$0.7x10$^{-3}$ magnitudes) using equation 7 of \citet{Zucker2007}, with gravity- and limb-darkening coefficients taken from \citet{Claret2011} for the {\it V}-band assuming T$_{\rm eff}$ = 3500~K, log~$g$ = 5 dex, Z $=-$0.5 dex.
The beaming effect is caused by Doppler shifts in the spectrum and is determined by integrating the spectrum within a photometric bandpass \citep{Bloemen2011}.
We derive the beaming variability (8$\pm$3x10$^{-4}$ magnitudes) by integrating the HST/STIS spectrum inside the {\it V}-band passband (551$\pm$88nm) at both ends of the radial velocity amplitude shown in Figure~\ref{orbit} (240km/s). 
The reflected and thermal components are negligible for Wolf~1130 since the white dwarf is cool and the separation is still relatively large ($\sim$3~R$_{\odot}$).
The middle panel of Figure~\ref{phot_fits} shows the combined amplitude of variation for the calculated ellipsoidal and beaming effects.
The residual variation in the light curve, right panel of Figure~\ref{phot_fits}, is not caused by the orbit and is likely caused by long-lived starspots.
The coherence of the residual photometry, when phase folded to the orbital period, validates the tidal locking assumption.

\begin{sidewaysfigure*}
\epsscale{1}
\figurenum{8}
\centering
\includegraphics[width=5in,angle=90]{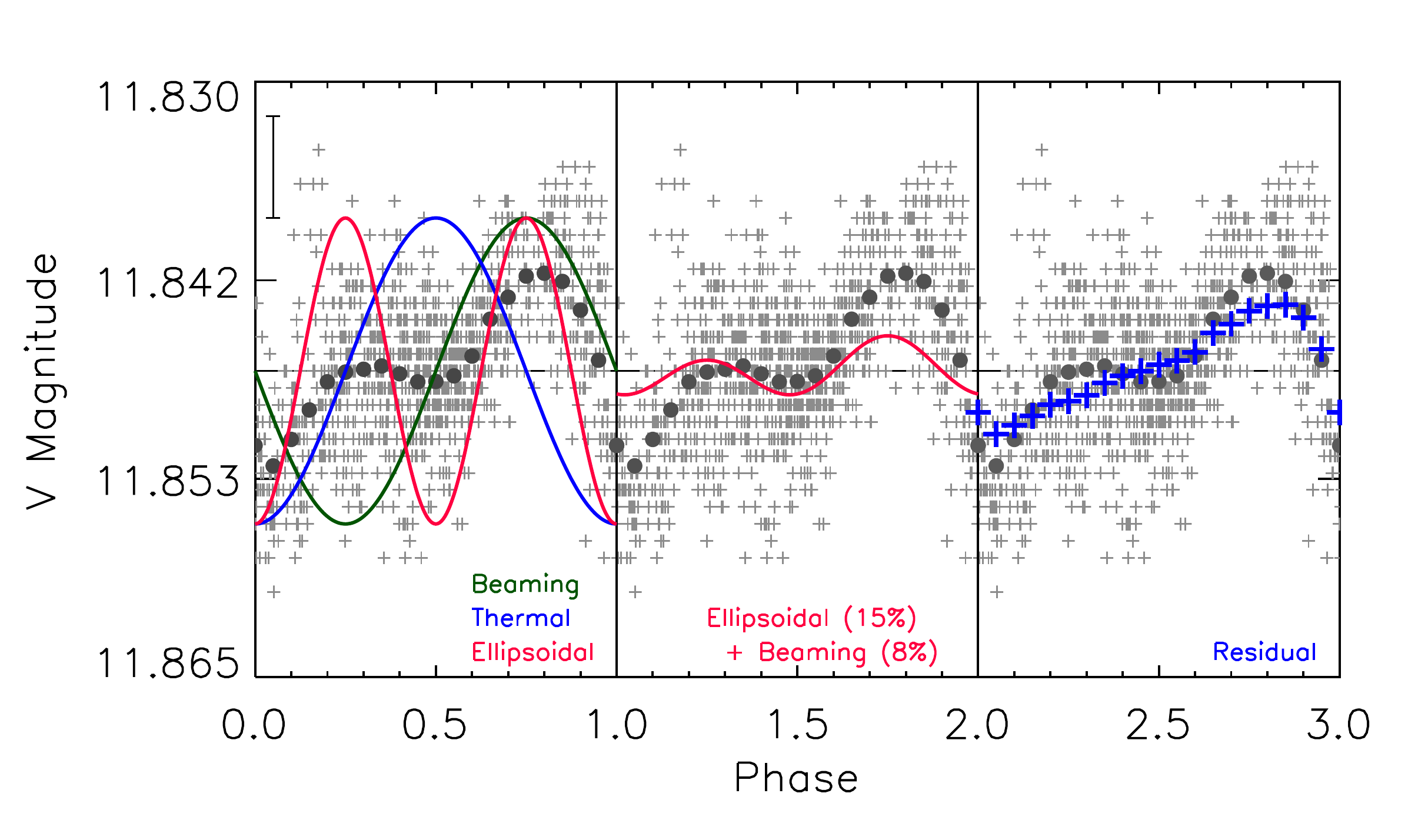}
\caption{Wolf~1130 {\it V}-band light curve from Figure~\ref{phot} with models of orbit induced variations. \emph{Left}: The phase dependence of tidal elongation (ellipsoidal), thermal (day/night), and Doppler beaming variations are unique, but none are good individual matches to the data. 
\emph{Middle}: The combination of calculated ellipsoidal and beaming amplitudes account for $\sim$20$\%$ of the observed variability.
\emph{Right}: The coherence of the photometry after removing orbit induced changes implies long-lived starspots and solidifies the assertion that Wolf~1130B is tidally locked.
\label{phot_fits}}
\end{sidewaysfigure*}

\item \textbf{Rotation - }
\citet{mann2016a} outlined the method used to determine rotational velocities in IGRINS spectra.
The highest signal-to-noise spectrum of Wolf~1130 was matched to a BT-SETTL \citep{allard2012,allard2014} model with similar parameters to the one shown in Figure~\ref{optical_spec} (T$_{\rm eff}$=3500~K, log~$g$=4.83, [Fe/H]$=-$1.3).
The synthetic spectrum was then artificially broadened with the IDL code lsf\_rotate \citep{gray1992,hubeny2011} and convolved with a Gaussian estimate of the instrumental broadening, which was determined from a simultaneous fit to telluric lines.
Values of $v$~sin~$i$ between 0-50~\kms\ were used to determine the best fit model via chi-squared minimization.
The measured $v$~sin~$i$ and uncertainties were taken to be the median and standard deviation across all 44 IGRINS orders.
For Wolf~1130A we measured $v$~sin~$i$=14.7 $\pm$ 0.7 \kms. This is consistent with \citet{stauffer1986} who measured $v$~sin~$i$=15 \kms, and similar to $v$~sin~$i$=12.7 derived by \citet{houdebine2010}. However, \citet{gizis1998} estimated $v$~sin~$i$=30$\pm$5~\kms\ and this resulted in a different interpretation of how Wolf~1130B evolved. 
For comparison, we also determined $v$~sin~$i$ for Gl~494 (Ross~458) from the IGRINS spectrum shown in Figure~\ref{IGRINS_comp}. 
Our measurement of 10.5$\pm$0.6~\kms\ is in good agreement with \citet{houdebine2010} who measure $v$~sin~$i$~$=$9.75 \kms. 
Additional comparison to the models in Figure~\ref{moog_ca} support our $v$~sin~$i\sim$15 \kms\ measurement for Wolf~1130A.
The $v$~sin~$i$ we measure is high for M dwarfs as old as Wolf~1130A \citep{newton2016} and the typical rotation period for an old and inactive M dwarf is $>$10 days \citep{newton2017}.

\item \textbf{Temperature and Gravity - }
From the MDM and STIS optical spectra of Wolf~1130A we derived an effective temperature of 3530$\pm$60K using the weighted mean from the model fit method described by \citet{mann2013a}. The best fit model was found using the BT-SETTL model grid \citep{allard2014}, where 100,000 linear combinations of three synthetic spectra were used to find the best fit linear combination. The final parameters of the best-fit model to the MDM spectrum were T$_{\rm eff}$ $=$ 3595K, log~$g$ = 4.9 dex, and [Fe/H]$=-$1.1 dex. For the STIS spectrum, we found T$_{\rm eff}$ = 3500~K, log~$g$ = 4.83 dex, and  [Fe/H] $=-$1.3 dex. We trust the temperature and gravity measurements from the model fits to the broad optical spectra because the method is well calibrated throughout the subdwarf sequence \citep{lepine2007}. 

\item \textbf{Metallicity - }
We calculated [Fe/H] for Wolf 1130A using the IRTF/SpeX spectrum and following the techniques outlined by \citet{mann2013b}. 
That paper presented empirical relations between strong atomic lines in near-infrared M dwarf spectra and overall metallicity. These relations were calibrated using wide binaries containing an FGK primary and an M dwarf companion under the assumption of identical metallicities between binary components. 
In this work we adopt the mean of the $H$- and $K$-band relations [Fe/H]$=-$0.70 $\pm$ 0.12 as the metallicity for Wolf 1130A. Uncertainties account for Poisson errors in the spectrum as well as the scatter in the \citet{mann2013b} calibrations. 

The metallicity values in the literature have a large scatter but are all significantly subsolar. 
As shown by \citet{neves2012}, most methods for determining metallicity diverge below $-$0.6~dex.
\citet{rojas2012} derive [M/H]$=-$0.45$\pm$0.12 and [Fe/H]$=-$0.64$\pm$0.17 from K-band indices. 
Other [Fe/H] calculations include $-$0.80 dex \citep{woolf2009}, $-$0.87 dex \citep{stauffer1986b}, $-$0.89 dex \citep{bonfils2005}, and $-$1.02 dex \citep{schlaufman2010}.
\citet{schmidt2009} compiled [Fe/H]$=-$0.62$\pm$0.10 and [Ti/H]$=-$0.22$\pm$0.09 abundances for Wolf~1130A from \citet{woolf2006}, who used the MOOG spectral synthesis software \citep{sneden1973}. Additionally, \citet{schmidt2009} derived [O/H]$=-$0.45$\pm$0.11 from TiO lines in R$\sim$30,000 optical spectra.
Overall, our methods are most similar to \citet{rojas2012} and \citet{schmidt2009} and we include their measurements in Table~\ref{tbl-4}.

Employing the current version of the MOOG synthesis software with a Python wrapper\footnote{\url{https://bitbucket.org/madamow/pymoogi.git}, written by Monika Adamow}, we measured abundances in the IGRINS infrared spectrum. 
For this synthesis we used the parameters already identified above (T$_{\rm eff}$=3600~K, log~$g$=5, [M/H]$=-$1, and microturbulance=1.0~\kms) to select similar atmosphere models. 
Our synthetic spectra employed \citet{kurucz1979,kurucz11} models, that were broadened to the instrument resolution and $v$~sin~$i$=15.0 \kms\ and then modulated for different atomic abundances.
Figure~\ref{moog_ca} shows the K-band Ca~I triplet along with synthetic abundances of [Ca/H]$=-$0.2$\pm$0.1 dex. 
The 10$\%$ depth of the Ca lines, relative to the continuum, are some of the deeper lines in the infrared spectrum of Wolf~1130A. 
H$_2$O lines throughout the H and K bands, weak metal lines, and a relatively large $v$~sin~$i$ $\sim$15 \kms\ complicate a detailed abundance analysis of all the species identified in Figure~\ref{IGRINS_spec1} and prohibits our own determination of [Fe/H] from the IGRINS spectrum.
For strong OH and Ca lines in the IGRINS spectrum we measure [Ca/H]$=-$0.20$\pm$0.05 and [O/H]$=-$0.5$\pm$0.1, consistent with \citet{schmidt2009}. 
The abundance ratios relative to Fe from the literature ([Ti/Fe]$=$0.48$\pm$0.15, [O/Fe]$=$0.15$\pm$0.16, [Ca/Fe]$=$0.5$\pm$0.13) are all consistent with the alpha-element enhancements of the thick disk population \citep[e.g.,][]{reddy2006,bensby14}.

\begin{figure}
\epsscale{0.9}
\figurenum{9}
\centering
\includegraphics[width=5in]{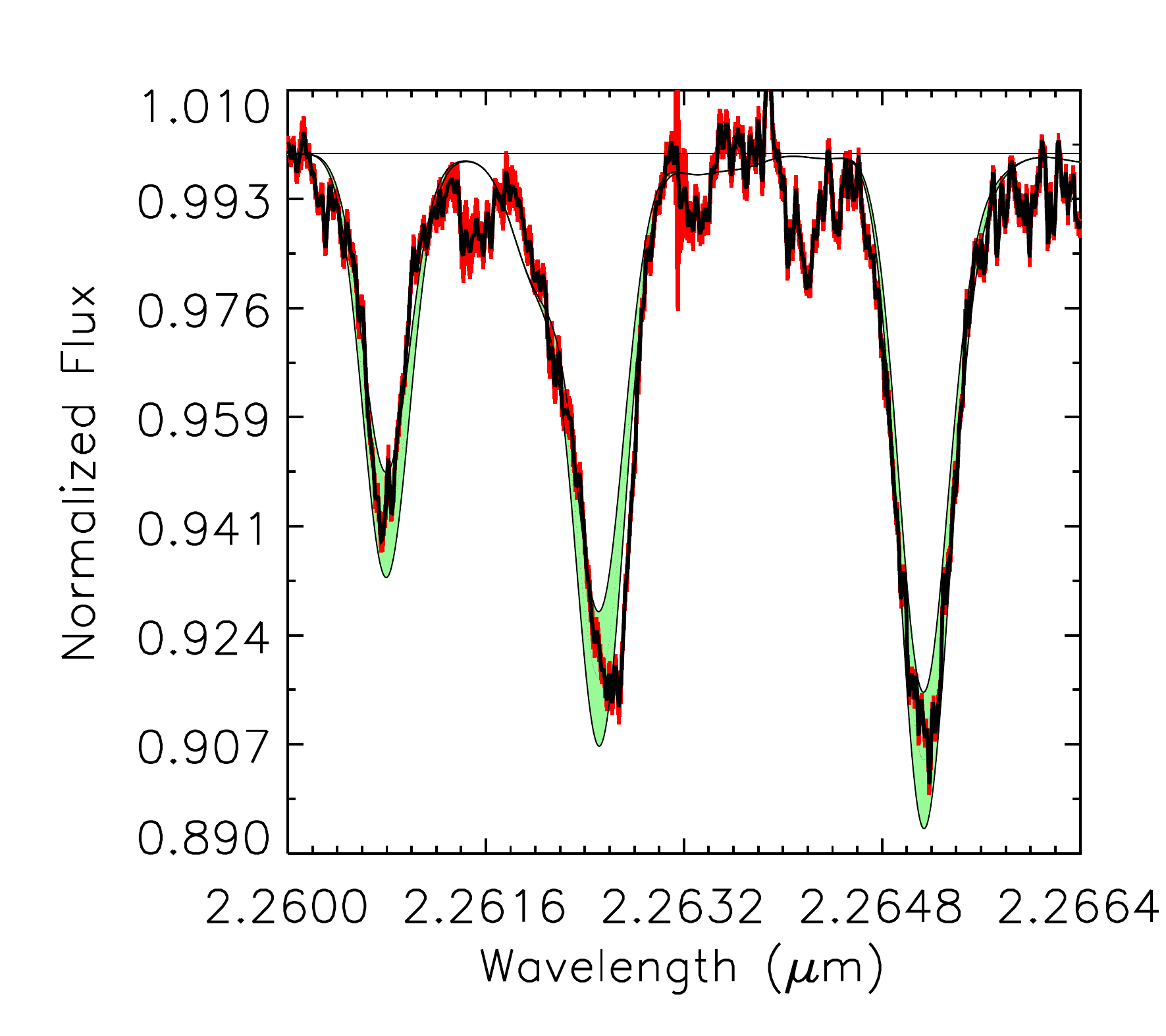}
\caption{The 40-epoch combined K-band spectrum of the Ca~I triplet in Wolf~1130A (black) with uncertainties (red). The Ca abundance is derived using the MOOG spectral synthesis software and \citet{kurucz1979,kurucz11} atmosphere models with T$_{\rm eff}$=3600~K, log~$g$=5, [M/H]$=-$1, and microturbulance=1.0~\kms. The green shaded area is for an abundance of [Ca/H]$=-$0.2$\pm$0.1 dex. The offset of the middle line is due to line location and strength deficiencies in the models. Additionally, this figure shows the agreement between our measured $v$~sin~$i$=14.7$\pm$0.7\kms\ and the 15\kms\ broadening of the model spectrum.
\label{moog_ca}}
\end{figure}

\item \textbf{Magnetic Field - }
The activity-age-rotation relation predicts that M dwarfs spin-down as they age and become less active \citep{douglas2014, newton2016}, but Wolf~1130A's rotation and activity is maintained by the tidal effects of its massive companion. 
With a rotation period of $\sim$12 hours, Wolf~1130A has an active chromosphere and we identify numerous emission lines (Figure~\ref{optical_emission}).
Since rotation and activity are presumed to be linked by the stellar magnetic field \citep{birkby2012,newton2017}, activity implies that strong fields could be present on Wolf~1130A.

In order to measure the magnetic field we employed a modified version of the MOOG spectral synthesis code, called MOOGStokes\footnote{\url{https://github.com/soylentdeen/MoogStokes}} \citep{deen2013}, which accounts for the Zeeman broadening of spectral lines.
Using the temperature, metallicity, $v$~sin~$i$ and surface gravity derived in the previous sections, we synthesized Zeeman broadened spectra from MARCS atmosphere models \citep{gustafsson2008}. 
Strong Na and Ti lines in the K-band are sensitive to magnetic field strength \citep{doppmann2003,Sokal2018} and we measure an upper limit on the magnetic field of 3~kG. 
The limitations to this measurement are the relatively high $v$~sin~$i$ and log~$g$ of Wolf~1130A, which broaden and weaken the lines. 
Above 3~kG the model Na and Ti lines are distinctly split, which is not seen in our high signal-to-noise IGRINS spectrum.
Numerous active M dwarfs have 2-4~kG magnetic fields \citep{johnskrull1996,johnskrull2000,shulyak2014} and Wolf~1130A does not have an exceptionally strong field.

\item \textbf{Radius - }
Using the {\it Gaia} DR1 parallax, we converted the 2MASS K$_s$ magnitude for Wolf~1130A to M$_K$=7.00$\pm$0.04 mag and, along with the SpeX-derived [Fe/H], apply the empirical M$_K$-Radius-[Fe/H] relationship from \citet{mann2015} to determine a radius of 0.302$\pm$0.009 R$_{\odot}$. For comparison, the Dartmouth Stellar Evolution Models \citep{dotter2008} produce a radius of 0.289$\pm$0.011 R$_{\odot}$. As noted by \citet{boyajian2012}, interior models are not fully calibrated and are generally $\sim$10$\%$ smaller than definitive radii from interferometric observations. 

\item \textbf{Mass - }
The absolute K-band magnitude also allows the application of the M$_K$-Mass relationship from \citet{benedict2016} to determine the mass of Wolf~1130A (0.308$\pm$0.016 M$_{\odot}$). The Dartmouth Stellar Evolution Models find a similar mass of 0.297$\pm$0.011 M$_{\odot}$. In our analysis we have chosen to fix the M subdwarf mass at 0.3~M$_{\odot}$.

\item \textbf{Rotational Inclination - }
Using the derived radius above and a rotation period equal to the orbital period (0.4967 days), we calculate an equatorial velocity of 30.3 $\pm$ 1.5 \kms\ and the rotation axis inclination of 29$\pm$2$^{\circ}$. As we discuss in Section 5, tidal locking ensures spin-orbit alignment on short timescales and so the orbital inclination can be assumed to equal the rotational inclination.

\item \textbf{Age - }
Ages of field M dwarfs from metallicity measurements are not precise \citep{reddy2006,newton2014}.
Age-Activity-Rotation relations for M dwarfs \citep{douglas2014,newton2017} break down in close binaries that artificially maintain high $v$~sin~$i$. 
As discussed in \citet{mace2013b}, the UVW velocities of Wolf 1130 are consistent with old disk-halo membership requirements from \citet{leggett1992}. 
The old, subdwarf nature of the M star Wolf~1130A is also supported by the TiO and CaH indices from optical spectra \citep{reid1995,lepine2013,gizis2016}. 
For a sample of F and G stars with thick-disk classifications, based on UVW velocities and metallicity like Wolf~1130A, \citet{reddy2006} determine an age of more than 10~Gyr.
We consider these kinematic properties to be the most reliable age limits for the Wolf~1130 system.

\end{itemize}

\subsection{Wolf 1130B - White Dwarf}

The combined visible-light and infrared orbital solution produces a minimum mass for Wolf~1130B (WD~2003+542) of 0.341~M$_{\odot}$. 
This mass is in agreement with what \citet{gizis1998} derived by assuming tidal locking and a model derived radius for Wolf~1130A of 0.28 R$_{\odot}$ \citep{gizis1997, baraffe1997}.
However, the $v$~sin~$i$ (30$\pm$5~\kms) measured by \citet{gizis1998} implies a nearly edge-on orbit and results in a low-mass (M$\sim$0.35M$_{\odot}$) He core white dwarf. 
However, we determine a mass for the primary of $1.24^{+0.19}_{-0.15}$ M$_{\odot}$ when we make the same assumptions as \citet{gizis1998}, with the exception of the smaller $v$~sin~$i$=14.7$\pm$0.7~\kms.

A model fit to the the STIS spectrum of Wolf~1130AB (Figure~\ref{optical_spec}) shows that there is no excess flux above the M dwarf continuum. 
The H~I absorption lines typical of white dwarf spectra \citep{wesemael1993} are not identified within the deep molecular absorption bands and H~I emission of Wolf~1130A.
Figure~\ref{optical_spec} includes curves for 7,000, 10,000, 15,000 and 20,000~K black bodies that are at the same distance as Wolf~1130 and have a radius of 0.005 R$_{\odot}$. 
Blueward of 3200~\AA, a Planck curve with a temperature greater than 7,000~K would be measurable. 
Additionally, Wolf~1130 was detected by the Galaxy Evolution Explorer \citep[GALEX,][]{Martin2005} in the near-ultraviolet (NUV, 1800-2750~\AA), and its absolute NUV magnitude is consistent with early-type M dwarfs in the compilation by \citet{Lepine2011}.

Based on our analysis of the M subdwarf in the previous section and the lack of flux from Wolf~1130B, we posit that the primary is most likely a massive white dwarf that has cooled to $<$7000~K.
However, a neutron star at any temperature would remain unseen since its radius would be two orders of magnitude smaller than the assumed white dwarf radius.
A white dwarf that is $1.24^{+0.19}_{-0.15}$ M$_{\odot}$ may have formed from a single progenitor, or from a merger of two white dwarfs.
If it formed from the merger of two $\sim$0.6~M$_{\odot}$ CO white dwarfs, then it would be composed of CO despite its high mass.
It is convenient that 0.6~M$_{\odot}$ white dwarfs are the most abundant \citep{reid1996,kepler2016}.
However, the stability of the close M subdwarf through the post-main-sequence evolution and merger of two white dwarfs would be difficult, if not impossible, to sustain. 
Additionally, there is a non-zero probability that this system is not primordial and formed through dynamical interactions in a cluster environment \citep{Kouwenhoven2010}.

The most likely pathway of evolution for Wolf~1130B was with a single progenitor, which would have required an initial mass between 6-8 M$_{\odot}$ and result in an ONe white dwarf \citep{Catalan2008,Cummings2016}. This single progenitor would have spent 50-100~Myr on the main sequence \citep{wood1992, monteiro2006}.
The white dwarf cooling age for Wolf~1130B depends on its composition and its temperature.
An ONe white dwarf with T$_{\rm eff}$ $<$7000 K and a mass of $1.24^{+0.19}_{-0.15}$ M$_{\odot}$ is at least $\sim$3.4 Gyr old \citep{althaus2007}. 
If the progenitor was 6-8~M$_{\odot}$, then its main-sequence lifetime is negligible and the lower age limit for Wolf~1130B is essentially just the 3.4~Gyr cooling age \citep{gilpons2001}.
Without directly detecting the massive primary and determining its temperature or composition, we cannot employ it to place significant bounds on the age of the Wolf~1130 system.

\subsection{Wolf 1130C - T8 Subdwarf Brown Dwarf}

The T8 subdwarf brown dwarf discovered by \citet{mace2013b} has a unique spectral morphology indicative of low metallicity and high surface gravity (Logsdon et al. submitted). 
For an age of 10~Gyr, \citet{baraffe2003} estimated the mass of an 800~K brown dwarf to be 0.050~M$_{\odot}$ and Wolf~1130C may be older and more massive than this.
These evolutionary tracks show that Wolf~1130C would have formed as a $\sim$2800~K late-type M dwarf. 
Combining evolutionary tracks with spectral type and temperature relationships \citep{filippazzo2015} reveals that Wolf~1130 would have become an L dwarf after $\sim$300~Myr and then a T dwarf after an additional 1.5~Gyr. 

The new distance to Wolf~1130 derived from the {\it Gaia} DR1 parallax is $\sim$5$\%$ farther than what was determined from the Hipparcos parallax and changes the calculated separation of Wolf~1130C to 3150$\pm$40AU.
Absolute magnitudes are also slightly brighter, M$_H$=18.46$\pm$0.10 and M$_{W2}$=13.85$\pm$0.08, but within the uncertainties presented by \citet{mace2013b}. These new absolute magnitudes maintain the subdwarf classification for Wolf~1130C, which sits below the trend line set by the rest of the T dwarf population. 

Table~\ref{tbl-5} lists the T dwarfs in multiple systems tabulated by \citet{deacon2017}.
Each of the five known systems is unique. Gl~570D \citep{burgasser2000} is the closest multiple to Earth with a wide separation T dwarf, making it a relatively bright benchmark.
Ross~458C \citep{goldman2010, burgasser2010, scholz2010} is the youngest and Wolf~1130C \citep{mace2013b} is the oldest, but they have similar spectral types and effective temperatures ($\sim$800~K).
$\xi$~UMa~E \citep{wright2013} is at the widest separation from the hierarchical binaries at the center of the system.
2MASS~J0213+3648~C \citep{deacon2017} stands out at the most distant, smallest separation, and warmest T dwarf companion in a multiple system.

There are a number of possible trends to note in the small sample of T dwarfs in multiple systems. 
First, the surveys that enabled these T dwarf discoveries (primarily 2MASS \citep{skrutskie2006}, UKIDSS \citep{lawrence2007}, and WISE \citep{wright2010}) have probed successively farther from the Sun.
Next, the separation between the T dwarf and the primary of the system may go up as a function of the primary star mass.
Wolf~1130 is the second most massive system after $\xi$UMa, consistent with the second largest separation. 
However, when Wolf~1130B was a 6-8~M$_{\odot}$ star it would have outweighed all of the other systems, and as discussed by \citet{dayjones2011b} the separation would have been a factor of $\sim$8 smaller (M$_{MS}$/M$_{WD}$ $\sim$ 8; \citet{jeans1924,zuckerman1987}), $\sim$400~AU.
\citet{bate2009} shows that formation models of multiples preferentially form extreme-mass-ratio systems with separations between 50 and 1200~AU.
Given the mass loss history of Wolf~1130B, it is likely that Wolf~1130C formed at a more typical separation before moving out to its current orbit of $\sim$3150~AU.
However, formation and cluster interactions may have been a less orderly process that resulted in the hardened inner binary and distant tertiary we see today \citep{reipurth2012}.
Additionally, we cannot rule out the capture of Wolf~1130C by Wolf~1130AB in the period of star forming cluster dissolution \citep{Kouwenhoven2010}. 
In all these scenarios, the system is coeval to within a few Myr and we can safely assume that the properties of Wolf~1130C match Wolf~1130AB.

\section{Discussion}

\subsection{Assumptions}

The single largest assumption we have made is that the system is tidally locked, which directly impacts the inclination we have determined.
The timescales for synchronization and circularization are strongly dependent on the orbital period of the binary \citep{zahn1975,zahn1977,zahn1989}.
For the short $\sim$12 hour period of Wolf~1130AB the circularization timescale is $<$~50,000 years and the synchronization timescale is an order of magnitude smaller \citep{hilditch2001}.
Because synchronization occurs more rapidly than circularization \citep{claret1995}, the zero eccentricity of the combined orbital solution implies spin-orbit alignment in the system. 
Additionally, the coherence of the phase-folded {\it V}-band photometry and the residual flux in Figure~\ref{phot_fits} provides strong evidence for tidal locking.

The radius and mass of Wolf~1130A derived from empirical relationships \citep{mann2015,benedict2016} are assumed to be the most reliable we have.
Increasing the radius by 10$\%$, perhaps to account for interior magnetic effects induced by tidal locking \citep{feiden2012}, would decrease the inclination by 10$\%$ (to 26$^{\circ}$) and would increase the mass of the white dwarf to well above the Chandrasekhar limit ($\sim$1.55 M$_{\odot}$). 
Since we do not measure a strong magnetic field, and the spot modulation is small amplitude, a substantially inflated radius for Wolf~1130A is unlikely.

In this work we have combined parameter calibrations from BT-SETTL \citep{allard2012,allard2014,mann2013a,mann2016a}, Kurucz \citep{kurucz1979,kurucz11,sneden1973}, MARCS \citep{gustafsson2008, deen2013}, and the Dartmouth Stellar Evolution Models \citep{dotter2008}. 
This is partly because no single model grid covers the parameter space that we require, partly because empirical relationships have been calibrated against specific models, and partly to show that the parameters derived from different methods are consistent in describing Wolf~1130.

At the youngest age we estimate for Wolf~1130B (3.4~Gyr, Section~4.2), the ONe white dwarf would be just below our detection limits. 
The M subdwarf (Wolf~1130A) metallicity and UVW velocities imply an age $>$10~Gyr \citep{reddy2006}. 
Additionally, \citet{rojas2012} and \citet{muirhead2013} identify a metallicity enrichment of up to +0.4~dex in post-common-envelope M dwarfs relative to typical field M dwarfs \citep{muirhead2012}.
Adjusting the metallicity-based age estimate of Wolf~1130A to account for contamination has the potential of adding Gyr to the system age.
We consider the $>$10~Gyr kinematic age be the most reliable for the Wolf~1130 system.

\subsection{Applications}

The best use of Wolf~1130, and other multiples containing T dwarfs, will be to measure the intersection of various stellar and substellar populations in model parameter space.
The typical separation of brown dwarfs in binaries is small \citep[around 3~AU, ][]{burgasser2007,bardalez2014,prato2015} and in multiple-body systems the likelihood of breakup through dynamic evolution is high \citep{reipurth2015}.
It is possible that Wolf~1130C has been excluded from the common envelope and escaped contamination, preserving the primordial metallicity of the entire system.
Determining the metallicity of the T dwarf directly will be difficult since it is faint (J$=$19.6 mag) and sub-solar metallicity atmosphere models are not calibrated for T dwarfs \citep[][Logsdon et al. submitted]{martin2017}.
Instruments like the Giant Magellan Telescope Near-Infrared Spectrograph \citep{jaffe2016}, in concert with atmospheric retrieval methods that maximize the utility of model grids \citep{line2017}, may facilitate the eventual characterization of Wolf~1130C.
For now, the abundances determined from Wolf~1130A provide limits on the metallicities of Wolf~1130C and motivate lower-metallicity model development (Logsdon et al. submitted).

While theoretical calculations hint at an overall multiplicity rate for the star formation process of $\sim$10$\%$ \citep{bate2009}, the number of known T dwarfs in multiples is only $\sim$1$\%$.
This implies that a unique evolutionary path is required in order to keep substellar companions bound to stellar binaries after formation and cluster dissolution. 
Brown dwarfs that form through the fragmentation of massive circumstellar disks \citep{bonnell1994,kratter2006,stamatellos2011} would reside in high-mass-ratio systems \citep{bate2003,bate2009}.
Yet, the formation of multiple systems through cloud fragmentation \citep{reipurth2012} and subsequent capture \citep{Kouwenhoven2010} will allow for more random system properties, which are testable by observation.
The direct application of formation models to reproduce the wide-separation T dwarfs in Table~\ref{tbl-5}, discovering a larger sample of T dwarfs in multiple systems at intermediate separations, and the identification and characterization of field subdwarf brown dwarfs like Wolf~1130C and the subdwarf T6 discovered by \citep{burningham2014} would each improve our understanding of the small occurrence rate of T dwarfs in multiple systems. 

A white dwarf as massive as Wolf~1130B is rare \citep{kepler2016} and exotic system evolution scenarios are plausible.
The upper limit of the mass we determine for Wolf~1130B is 1.43~M$_{\odot}$ and a neutron star primary in this system could have formed through the merger of a massive white dwarf and an M dwarf in the common-envelope phase of the system.
Any earlier merger activity would have increased the angular momentum and slowed tidal locking, requiring that the system be old enough to relax again.
Additionally, Wolf~1130 could be the result of a dynamical capture between an M and T dwarf binary and massive white dwarf or its progenitor.
Yet, strong interactions with other stars that would harden the inner binary are unlikely \citep{bonnell2002} and in-situ formation and evolution \citep{bate2009} is the most probable origin of Wolf~1130.

\subsection{What Happens to a Cataclysm Deferred?}


Wolf~1130A does not fill its Roche lobe (0.82~R$_{\odot}$) and it can only achieve this by moving closer to Wolf~1130B or by increasing its radius.
The main-sequence lifetime of a 0.3M$_{\odot}$ star is $\sim$200 Gyr, which is the longest timescale we should consider for this system's evolution.
The decreased separation of Wolf~1130AB through angular momentum loss will reduce the Roche lobe radius \citep[see reviews on cataclysmic variables by][]{warner1995,hellier2001} until mass transfer commences. 
At that point, Wolf~1130AB will become a cataclysmic variable and the separation will only be $\sim$1.1R$_{\odot}$ (assuming no tidal elongation, which would actually be $\sim$30$\%$ of the M dwarf radius once the separation is this small \citep{fitzpatrick2012}).
The primary mechanism for angular momentum loss in a cataclysmic variable is magnetic braking when orbital periods are longer than a few hours, and gravitational radiation (waves) at the shortest periods \citep{iben1985,iben1993,hellier2001,schreiber2003}.

Magnetic braking is not a rapid process since it relies on the stellar wind and magnetic field interactions to slowly sap the angular momentum. 
Employing the equations for magnetic wind braking \citep{sills2000,andronov2003} from \citet{muirhead2013} (who assumes efficient spin-orbit coupling) we find that Roche lobe overflow will start $\sim$6.2~Gyr from today.
Uncertainties in this age estimate are likely on the order of Gyr because the magnetic pole alignment of the M and white dwarfs create complicated fields that directly impact the rate of angular momentum loss \citep{wu1993,wheeler2012}.

Once mass transfer is initiated, magnetic braking will gain strength as material crosses magnetic field lines to form an accretion disk around the white dwarf \citep{hellier2001}. Since Wolf~1130A will be only $\sim$1.1~R$_{\odot}$ from Wolf~1130B, the magnetic braking and gravitational radiation timescales will be about the same order of magnitude and the cataclysmic variable phase will last $<$500 Myr \citep{kolb1996}. Tidal locking and alignment of the magnetic fields between the white dwarf and subdwarf may result in a sufficiently high mass transfer rate to lead to a Type~Ia supernova \citep{wheeler2012}, even if the white dwarf is ONe rather than CO \citep{marquardt2015}, or below the Chandrasekhar limit \citep{dessart2014, scalzo2014}. The low range of the combined Wolf~1130AB system mass (1.39~M$_{\odot}$) is significantly above the mass of sub-luminous, SN~1991bg-like Type~Ia events \citep{filippenko1992, Ruiz1993, blondin2017}. The age ($\gtrsim 3.5$~Gyr) and inspiral time ($\sim 6$~Gyr) of this system are consistent with the class of long-delay Type~Ia supernovae \citep{maoz2010}.
This is the first ultramassive white dwarf within 25~pc \citep{holberg2016, toonen2017} and the proximity of this system makes it an ideal candidate for follow-up studies and modeling of potential Type~Ia progenitors.

\section{Summary}
Wolf~1130AB is a nearby (16.7$\pm$0.2~pc) and old ($>$10~Gyr) pre-cataclysmic variable with a 0.4967 day orbital period. 
The combination of archival data with new optical and infrared observations between 0.2 and 2.5$\mu$m produces properties for each component in the system.
The M subdwarf (Wolf~1130A) is the dominant flux source in the system but not the most massive (M$_{A}$ $\sim$0.3~M$_{\odot}$). 
Wolf~1130A is tidally locked, metal-poor ([Fe/H]$=-$0.7$\pm$0.12 dex), and shows low-level photospheric variability indicative of spots.
The ultramassive ($1.24^{+0.19}_{-0.15}$~M$_{\odot}$) ONe white dwarf component in the system (Wolf~1130B) remains unseen.
Without a direct detection of the massive component in this system, we can't precisely establish its age and future observations at UV and X-ray wavelengths may reveal its nature.
This is the first known system containing a potential Type~Ia progenitor, with a mass near or above the Chandrasekhar limit, within 25~pc \citep{holberg2016, toonen2017}.

A distant ($\sim$3150~AU) T subdwarf component of the system (Wolf~1130C) shows a spectral morphology consistent with old age and high-mass \citep{mace2013b}.
Wolf~1130C is distinctly on the edge of model parameter space with the lowest metallicity, a small radius, high mass and large surface gravity (\citet{mace2013b}, Logsdon et al. submitted) and is the oldest of only five known T dwarfs in multiple systems \citep{deacon2017}.
The faintness of the T dwarf and white dwarf in this system limit their utility in determining system parameters, but this can be overcome with future instrumentation \citep[GMTNIRS,][]{jaffe2016} and atmospheric retrieval methods \citep{line2017}.
The photometric precision of the The Transiting Exoplanet Survey Satellite (TESS) will be $<$0.1\% for Wolf~1130 \citep{ricker2014,stassun2017} and a fit to the light curve may yield more precise system parameters.
The dynamical evolution and initial mass ratio of Wolf~1130 need to be considered when modeling stellar interactions in the epoch of star formation and post-main-sequence evolution \citep{bate2009,gosnell2015}. 
The Wolf~1130 system is a unique test case for modeling star formation, dynamical evolution, post-main-sequence evolution, and white dwarf and brown dwarf cooling models.

\acknowledgments
We acknowledge the poem \emph{Harlem} by Langston Hughes, which provides an analogy for pre-cataclysmic variables like Wolf~1130AB. It reads, `What happens to a dream deferred? Does it dry up like a raisin in the sun? Or fester like a sore$-$ and then run? Does it stink like rotten meat? Or crust and sugar over$-$ like a syrupy sweet? Maybe it just sags like a heavy load. Or does it explode?'

We appreciate the comments and recommendations of the anonymous referee, Sarah Logsdon, Emily Martin, Fabiola Campos, Kimberly Sokal, Casey Deen, Julie Skinner Manegold, Edward Robinson, Mike Montgomery and Craig Wheeler. Thank you to Kevin Gullikson, Kyle Kaplan and Jacob McLane for exchanging observing time for this project.

This paper includes data taken at the McDonald Observatory of the University of Texas at Austin and we thank the Observer Support for their constant watch over the facility. Support was provided in part by the US National Science Foundation under grant AST-1616040 to C.~Sneden.

This work used the Immersion Grating Infrared Spectrometer (IGRINS) that was developed under a collaboration between the University of Texas at Austin and the Korea Astronomy and Space Science Institute (KASI) with the financial support of the US National Science Foundation under grant AST-1229522, of the University of Texas at Austin, and of the Korean GMT Project of KASI.

This work has made use of data from the European Space Agency (ESA) mission {\it Gaia} (\url{https://www.cosmos.esa.int/gaia}), processed by the {\it Gaia} Data Processing and Analysis Consortium (DPAC, \url{https://www.cosmos.esa.int/web/gaia/dpac/consortium}). Funding for the DPAC has been provided by national institutions, in particular the institutions participating in the {\it Gaia} Multilateral Agreement.

\pagestyle{empty}

\bibliographystyle{apj}
\bibliography{bibidy}

\begin{thebibliography}{}
\expandafter\ifx\csname natexlab\endcsname\relax\def\natexlab#1{#1}\fi
\providecommand{\url}[1]{\href{#1}{#1}}

\bibitem[{{Allard}(2014)}]{allard2014}
{Allard}, F. 2014, in IAU Symposium, Vol. 299, Exploring the Formation and
  Evolution of Planetary Systems, ed. M.~{Booth}, B.~C. {Matthews}, \& J.~R.
  {Graham}, 271--272

\bibitem[{{Allard} {et~al.}(2012){Allard}, {Homeier}, \&
  {Freytag}}]{allard2012}
{Allard}, F., {Homeier}, D., \& {Freytag}, B. 2012, Philosophical Transactions
  of the Royal Society of London Series A, 370, 2765

\bibitem[{{Althaus} {et~al.}(2007){Althaus}, {Garc{\'{\i}}a-Berro}, {Isern},
  {C{\'o}rsico}, \& {Rohrmann}}]{althaus2007}
{Althaus}, L.~G., {Garc{\'{\i}}a-Berro}, E., {Isern}, J., {C{\'o}rsico}, A.~H.,
  \& {Rohrmann}, R.~D. 2007, \aap, 465, 249

\bibitem[{{Andronov} {et~al.}(2003){Andronov}, {Pinsonneault}, \&
  {Sills}}]{andronov2003}
{Andronov}, N., {Pinsonneault}, M., \& {Sills}, A. 2003, \apj, 582, 358

\bibitem[{{Baraffe} {et~al.}(1997){Baraffe}, {Chabrier}, {Allard}, \&
  {Hauschildt}}]{baraffe1997}
{Baraffe}, I., {Chabrier}, G., {Allard}, F., \& {Hauschildt}, P.~H. 1997, \aap,
  327, 1054

\bibitem[{{Baraffe} {et~al.}(2003){Baraffe}, {Chabrier}, {Barman}, {Allard}, \&
  {Hauschildt}}]{baraffe2003}
{Baraffe}, I., {Chabrier}, G., {Barman}, T.~S., {Allard}, F., \& {Hauschildt},
  P.~H. 2003, \aap, 402, 701

\bibitem[{{Bardalez Gagliuffi} {et~al.}(2014){Bardalez Gagliuffi}, {Burgasser},
  {Gelino}, {Looper}, {Nicholls}, {Schmidt}, {Cruz}, {West}, {Gizis}, \&
  {Metchev}}]{bardalez2014}
{Bardalez Gagliuffi}, D.~C., {Burgasser}, A.~J., {Gelino}, C.~R., {et~al.}
  2014, \apj, 794, 143

\bibitem[{{Bate}(2009)}]{bate2009}
{Bate}, M.~R. 2009, \mnras, 392, 590

\bibitem[{{Bate} {et~al.}(2002){Bate}, {Bonnell}, \& {Bromm}}]{bate2002}
{Bate}, M.~R., {Bonnell}, I.~A., \& {Bromm}, V. 2002, \mnras, 332, L65

\bibitem[{{Bate} {et~al.}(2003){Bate}, {Bonnell}, \& {Bromm}}]{bate2003}
---. 2003, \mnras, 339, 577

\bibitem[{{Benedict} {et~al.}(2016){Benedict}, {Henry}, {Franz}, {McArthur},
  {Wasserman}, {Jao}, {Cargile}, {Dieterich}, {Bradley}, {Nelan}, \&
  {Whipple}}]{benedict2016}
{Benedict}, G.~F., {Henry}, T.~J., {Franz}, O.~G., {et~al.} 2016, \aj, 152, 141

\bibitem[{{Bensby} {et~al.}(2014){Bensby}, {Feltzing}, \& {Oey}}]{bensby14}
{Bensby}, T., {Feltzing}, S., \& {Oey}, M.~S. 2014, \aap, 562, A71

\bibitem[{{Birkby} {et~al.}(2012){Birkby}, {Nefs}, {Hodgkin}, {Kov{\'a}cs},
  {Sip{\H o}cz}, {Pinfield}, {Snellen}, {Mislis}, {Murgas}, {Lodieu}, {de
  Mooij}, {Goulding}, {Cruz}, {Stoev}, {Cappetta}, {Palle}, {Barrado},
  {Saglia}, {Martin}, \& {Pavlenko}}]{birkby2012}
{Birkby}, J., {Nefs}, B., {Hodgkin}, S., {et~al.} 2012, \mnras, 426, 1507

\bibitem[{{Bloemen} {et~al.}(2011){Bloemen}, {Marsh}, {{\O}stensen},
  {Charpinet}, {Fontaine}, {Degroote}, {Heber}, {Kawaler}, {Aerts}, {Green},
  {Telting}, {Brassard}, {G{\"a}nsicke}, {Handler}, {Kurtz}, {Silvotti}, {Van
  Grootel}, {Lindberg}, {Pursimo}, {Wilson}, {Gilliland}, {Kjeldsen},
  {Christensen-Dalsgaard}, {Borucki}, {Koch}, {Jenkins}, \&
  {Klaus}}]{Bloemen2011}
{Bloemen}, S., {Marsh}, T.~R., {{\O}stensen}, R.~H., {et~al.} 2011, \mnras,
  410, 1787

\bibitem[{{Blondin} {et~al.}(2017){Blondin}, {Dessart}, {Hillier}, \&
  {Khokhlov}}]{blondin2017}
{Blondin}, S., {Dessart}, L., {Hillier}, D.~J., \& {Khokhlov}, A.~M. 2017,
  \mnras, 470, 157

\bibitem[{{Bonfils} {et~al.}(2005){Bonfils}, {Delfosse}, {Udry}, {Santos},
  {Forveille}, \& {S{\'e}gransan}}]{bonfils2005}
{Bonfils}, X., {Delfosse}, X., {Udry}, S., {et~al.} 2005, \aap, 442, 635

\bibitem[{{Bonnell} \& {Bate}(1994)}]{bonnell1994}
{Bonnell}, I.~A., \& {Bate}, M.~R. 1994, \mnras, 271, astro-ph/9411081

\bibitem[{{Bonnell} \& {Bate}(2002)}]{bonnell2002}
---. 2002, \mnras, 336, 659

\bibitem[{{Boyajian} {et~al.}(2012){Boyajian}, {von Braun}, {van Belle},
  {McAlister}, {ten Brummelaar}, {Kane}, {Muirhead}, {Jones}, {White},
  {Schaefer}, {Ciardi}, {Henry}, {L{\'o}pez-Morales}, {Ridgway}, {Gies}, {Jao},
  {Rojas-Ayala}, {Parks}, {Sturmann}, {Sturmann}, {Turner}, {Farrington},
  {Goldfinger}, \& {Berger}}]{boyajian2012}
{Boyajian}, T.~S., {von Braun}, K., {van Belle}, G., {et~al.} 2012, \apj, 757,
  112

\bibitem[{{Brandt} \& {Huang}(2015)}]{Brandt2015}
{Brandt}, T.~D., \& {Huang}, C.~X. 2015, \apj, 807, 24

\bibitem[{{Burgasser} {et~al.}(2007){Burgasser}, {Reid}, {Siegler}, {Close},
  {Allen}, {Lowrance}, \& {Gizis}}]{burgasser2007}
{Burgasser}, A.~J., {Reid}, I.~N., {Siegler}, N., {et~al.} 2007, Protostars and
  Planets V, 427

\bibitem[{{Burgasser} {et~al.}(2000){Burgasser}, {Kirkpatrick}, {Cutri},
  {McCallon}, {Kopan}, {Gizis}, {Liebert}, {Reid}, {Brown}, {Monet}, {Dahn},
  {Beichman}, \& {Skrutskie}}]{burgasser2000}
{Burgasser}, A.~J., {Kirkpatrick}, J.~D., {Cutri}, R.~M., {et~al.} 2000, \apjl,
  531, L57

\bibitem[{{Burgasser} {et~al.}(2010){Burgasser}, {Simcoe}, {Bochanski},
  {Saumon}, {Mamajek}, {Cushing}, {Marley}, {McMurtry}, {Pipher}, \&
  {Forrest}}]{burgasser2010}
{Burgasser}, A.~J., {Simcoe}, R.~A., {Bochanski}, J.~J., {et~al.} 2010, \apj,
  725, 1405

\bibitem[{{Burleigh} {et~al.}(2011){Burleigh}, {Steele}, {Dobbie}, {Farihi},
  {Napiwotzki}, {Maxted}, {Barstow}, {Jameson}, {Casewell}, {Gaensicke}, \&
  {Marsh}}]{burleigh2011}
{Burleigh}, M.~R., {Steele}, P.~R., {Dobbie}, P.~D., {et~al.} 2011, in American
  Institute of Physics Conference Series, Vol. 1331, American Institute of
  Physics Conference Series, ed. S.~{Schuh}, H.~{Drechsel}, \& U.~{Heber},
  262--270

\bibitem[{{Burningham} {et~al.}(2014){Burningham}, {Smith}, {Cardoso}, {Lucas},
  {Burgasser}, {Jones}, \& {Smart}}]{burningham2014}
{Burningham}, B., {Smith}, L., {Cardoso}, C.~V., {et~al.} 2014, \mnras, 440,
  359

\bibitem[{{Catal{\'a}n} {et~al.}(2008){Catal{\'a}n}, {Isern},
  {Garc{\'{\i}}a-Berro}, \& {Ribas}}]{Catalan2008}
{Catal{\'a}n}, S., {Isern}, J., {Garc{\'{\i}}a-Berro}, E., \& {Ribas}, I. 2008,
  \mnras, 387, 1693

\bibitem[{{Chubak} {et~al.}(2012){Chubak}, {Marcy}, {Fischer}, {Howard},
  {Isaacson}, {Johnson}, \& {Wright}}]{chubak2012}
{Chubak}, C., {Marcy}, G., {Fischer}, D.~A., {et~al.} 2012, ArXiv e-prints,
  arXiv:1207.6212

\bibitem[{{Claret} \& {Bloemen}(2011)}]{Claret2011}
{Claret}, A., \& {Bloemen}, S. 2011, \aap, 529, A75

\bibitem[{{Claret} {et~al.}(1995){Claret}, {Gimenez}, \& {Cunha}}]{claret1995}
{Claret}, A., {Gimenez}, A., \& {Cunha}, N.~C.~S. 1995, \aap, 299, 724

\bibitem[{{Crockett} {et~al.}(2012){Crockett}, {Mahmud}, {Prato},
  {Johns-Krull}, {Jaffe}, {Hartigan}, \& {Beichman}}]{crockett2012}
{Crockett}, C.~J., {Mahmud}, N.~I., {Prato}, L., {et~al.} 2012, \apj, 761, 164

\bibitem[{{Cummings} {et~al.}(2016){Cummings}, {Kalirai}, {Tremblay},
  {Ramirez-Ruiz}, \& {Bergeron}}]{Cummings2016}
{Cummings}, J.~D., {Kalirai}, J.~S., {Tremblay}, P.-E., {Ramirez-Ruiz}, E., \&
  {Bergeron}, P. 2016, \apjl, 820, L18

\bibitem[{{Cushing} {et~al.}(2004){Cushing}, {Vacca}, \&
  {Rayner}}]{cushing2004}
{Cushing}, M.~C., {Vacca}, W.~D., \& {Rayner}, J.~T. 2004, \pasp, 116, 362

\bibitem[{{Day-Jones} {et~al.}(2011){Day-Jones}, {Pinfield}, {Ruiz},
  {Beaumont}, {Burningham}, {Gallardo}, {Gianninas}, {Bergeron}, {Napiwotzki},
  {Jenkins}, {Zhang}, {Murray}, {Catal{\'a}n}, \& {Gomes}}]{dayjones2011b}
{Day-Jones}, A.~C., {Pinfield}, D.~J., {Ruiz}, M.~T., {et~al.} 2011, \mnras,
  410, 705

\bibitem[{{De Rosa} {et~al.}(2014){De Rosa}, {Patience}, {Ward-Duong}, {Vigan},
  {Marois}, {Song}, {Macintosh}, {Graham}, {Doyon}, {Bessell}, {Lai},
  {McCarthy}, \& {Kulesa}}]{derosa2014}
{De Rosa}, R.~J., {Patience}, J., {Ward-Duong}, K., {et~al.} 2014, \mnras, 445,
  3694

\bibitem[{{Deacon} {et~al.}(2017){Deacon}, {Magnier}, {Liu}, {Schlieder},
  {Aller}, {Best}, {Bowler}, {Burgett}, {Chambers}, {Draper}, {Flewelling},
  {Hodapp}, {Kaiser}, {Metcalfe}, {Sweeney}, {Wainscoat}, \&
  {Waters}}]{deacon2017}
{Deacon}, N.~R., {Magnier}, E.~A., {Liu}, M.~C., {et~al.} 2017, \mnras, 467,
  1126

\bibitem[{{Deen}(2013)}]{deen2013}
{Deen}, C.~P. 2013, \aj, 146, 51

\bibitem[{{Dessart} {et~al.}(2014){Dessart}, {Blondin}, {Hillier}, \&
  {Khokhlov}}]{dessart2014}
{Dessart}, L., {Blondin}, S., {Hillier}, D.~J., \& {Khokhlov}, A. 2014, \mnras,
  441, 532

\bibitem[{{Doppmann} \& {Jaffe}(2003)}]{doppmann2003}
{Doppmann}, G.~W., \& {Jaffe}, D.~T. 2003, \aj, 126, 3030

\bibitem[{{Dotter} {et~al.}(2008){Dotter}, {Chaboyer}, {Jevremovi{\'c}},
  {Kostov}, {Baron}, \& {Ferguson}}]{dotter2008}
{Dotter}, A., {Chaboyer}, B., {Jevremovi{\'c}}, D., {et~al.} 2008, \apjs, 178,
  89

\bibitem[{{Douglas} {et~al.}(2014){Douglas}, {Ag{\"u}eros}, {Covey}, {Bowsher},
  {Bochanski}, {Cargile}, {Kraus}, {Law}, {Lemonias}, {Arce}, {Fierroz}, \&
  {Kundert}}]{douglas2014}
{Douglas}, S.~T., {Ag{\"u}eros}, M.~A., {Covey}, K.~R., {et~al.} 2014, \apj,
  795, 161

\bibitem[{{Feiden} \& {Chaboyer}(2012)}]{feiden2012}
{Feiden}, G.~A., \& {Chaboyer}, B. 2012, \apj, 757, 42

\bibitem[{{Filippazzo} {et~al.}(2015){Filippazzo}, {Rice}, {Faherty}, {Cruz},
  {Van Gordon}, \& {Looper}}]{filippazzo2015}
{Filippazzo}, J.~C., {Rice}, E.~L., {Faherty}, J., {et~al.} 2015, \apj, 810,
  158

\bibitem[{{Filippenko} {et~al.}(1992){Filippenko}, {Richmond}, {Branch},
  {Gaskell}, {Herbst}, {Ford}, {Treffers}, {Matheson}, {Ho}, {Dey}, {Sargent},
  {Small}, \& {van Breugel}}]{filippenko1992}
{Filippenko}, A.~V., {Richmond}, M.~W., {Branch}, D., {et~al.} 1992, \aj, 104,
  1543

\bibitem[{{Fitzpatrick}(2012)}]{fitzpatrick2012}
{Fitzpatrick}, R. 2012, {An Introduction to Celestial Mechanics}

\bibitem[{{Frey} {et~al.}(2006){Frey}, {Leviton}, \& {Madison}}]{frey2006}
{Frey}, B.~J., {Leviton}, D.~B., \& {Madison}, T.~J. 2006, in \procspie, Vol.
  6273, Society of Photo-Optical Instrumentation Engineers (SPIE) Conference
  Series, 62732J

\bibitem[{{Gaidos} {et~al.}(2014){Gaidos}, {Mann}, {L{\'e}pine}, {Buccino},
  {James}, {Ansdell}, {Petrucci}, {Mauas}, \& {Hilton}}]{gaidos2014}
{Gaidos}, E., {Mann}, A.~W., {L{\'e}pine}, S., {et~al.} 2014, \mnras, 443, 2561

\bibitem[{{Garc{\'{\i}}a} \& {Mermilliod}(2001)}]{garcia2001}
{Garc{\'{\i}}a}, B., \& {Mermilliod}, J.~C. 2001, \aap, 368, 122

\bibitem[{{Gil-Pons} \& {Garc{\'{\i}}a-Berro}(2001)}]{gilpons2001}
{Gil-Pons}, P., \& {Garc{\'{\i}}a-Berro}, E. 2001, \aap, 375, 87

\bibitem[{{Gizis}(1997)}]{gizis1997}
{Gizis}, J.~E. 1997, \aj, 113, 806

\bibitem[{{Gizis}(1998)}]{gizis1998}
---. 1998, \aj, 115, 2053

\bibitem[{{Gizis} {et~al.}(2016){Gizis}, {Marks}, \& {Hauschildt}}]{gizis2016}
{Gizis}, J.~E., {Marks}, Z., \& {Hauschildt}, P.~H. 2016, \mnras, 455, 3824

\bibitem[{{Goldman} {et~al.}(2010){Goldman}, {Marsat}, {Henning}, {Clemens}, \&
  {Greiner}}]{goldman2010}
{Goldman}, B., {Marsat}, S., {Henning}, T., {Clemens}, C., \& {Greiner}, J.
  2010, \mnras, 405, 1140

\bibitem[{{Gosnell} {et~al.}(2015){Gosnell}, {Mathieu}, {Geller}, {Sills},
  {Leigh}, \& {Knigge}}]{gosnell2015}
{Gosnell}, N.~M., {Mathieu}, R.~D., {Geller}, A.~M., {et~al.} 2015, \apj, 814,
  163

\bibitem[{{Gray}(1992)}]{gray1992}
{Gray}, D.~F. 1992, {The observation and analysis of stellar photospheres.}

\bibitem[{{Gregg} {et~al.}(2006){Gregg}, {Silva}, {Rayner}, {Worthey},
  {Valdes}, {Pickles}, {Rose}, {Carney}, \& {Vacca}}]{gregg2006}
{Gregg}, M.~D., {Silva}, D., {Rayner}, J., {et~al.} 2006, in The 2005 HST
  Calibration Workshop: Hubble After the Transition to Two-Gyro Mode, ed. A.~M.
  {Koekemoer}, P.~{Goudfrooij}, \& L.~L. {Dressel}, 209

\bibitem[{{Gullikson} {et~al.}(2016){Gullikson}, {Kraus}, \&
  {Dodson-Robinson}}]{gullikson2016b}
{Gullikson}, K., {Kraus}, A., \& {Dodson-Robinson}, S. 2016, \aj, 152, 40

\bibitem[{{Gully-Santiago} {et~al.}(2012){Gully-Santiago}, {Wang}, {Deen}, \&
  {Jaffe}}]{gully2012}
{Gully-Santiago}, M., {Wang}, W., {Deen}, C., \& {Jaffe}, D. 2012, in
  \procspie, Vol. 8450, Modern Technologies in Space- and Ground-based
  Telescopes and Instrumentation II, 84502S

\bibitem[{{Gustafsson} {et~al.}(2008){Gustafsson}, {Edvardsson}, {Eriksson},
  {J{\o}rgensen}, {Nordlund}, \& {Plez}}]{gustafsson2008}
{Gustafsson}, B., {Edvardsson}, B., {Eriksson}, K., {et~al.} 2008, \aap, 486,
  951

\bibitem[{{Harrington} \& {Dahn}(1980)}]{harrington1980}
{Harrington}, R.~S., \& {Dahn}, C.~C. 1980, \aj, 85, 454

\bibitem[{{Hartkopf} \& {Mason}(2004)}]{Hartkopf2004}
{Hartkopf}, W.~I., \& {Mason}, B.~D. 2004, in Revista Mexicana de Astronomia y
  Astrofisica, vol.~27, Vol.~21, Revista Mexicana de Astronomia y Astrofisica
  Conference Series, ed. C.~{Allen} \& C.~{Scarfe}, 83--90

\bibitem[{{Hellier}(2001)}]{hellier2001}
{Hellier}, C. 2001, {Cataclysmic Variable Stars}

\bibitem[{{Hilditch}(2001)}]{hilditch2001}
{Hilditch}, R.~W. 2001, {An Introduction to Close Binary Stars}, 392

\bibitem[{{Hinkle} {et~al.}(1995){Hinkle}, {Wallace}, \&
  {Livingston}}]{hinkle1995}
{Hinkle}, K., {Wallace}, L., \& {Livingston}, W. 1995, \pasp, 107, 1042

\bibitem[{{Holberg} {et~al.}(2016){Holberg}, {Oswalt}, {Sion}, \&
  {McCook}}]{holberg2016}
{Holberg}, J.~B., {Oswalt}, T.~D., {Sion}, E.~M., \& {McCook}, G.~P. 2016,
  \mnras, 462, 2295

\bibitem[{{Horne}(1986)}]{horne1986}
{Horne}, K. 1986, \pasp, 98, 609

\bibitem[{{Houdebine}(2010)}]{houdebine2010}
{Houdebine}, E.~R. 2010, \mnras, 407, 1657

\bibitem[{{Hubeny} \& {Lanz}(2011)}]{hubeny2011}
{Hubeny}, I., \& {Lanz}, T. 2011, {Synspec: General Spectrum Synthesis
  Program}, Astrophysics Source Code Library, , , ascl:1109.022

\bibitem[{{Iben} \& {Livio}(1993)}]{iben1993}
{Iben}, Jr., I., \& {Livio}, M. 1993, \pasp, 105, 1373

\bibitem[{{Iben} \& {Tutukov}(1985)}]{iben1985}
{Iben}, Jr., I., \& {Tutukov}, A.~V. 1985, \apjs, 58, 661

\bibitem[{{{\it Gaia} Collaboration} {et~al.}(2016{\natexlab{a}}){{\it Gaia}
  Collaboration}, {Prusti}, {de Bruijne}, {Brown}, {Vallenari}, {Babusiaux},
  {Bailer-Jones}, {Bastian}, {Biermann}, {Evans}, \& et~al.}]{gaia2016}
{{\it Gaia} Collaboration}, {Prusti}, T., {de Bruijne}, J.~H.~J., {et~al.}
  2016{\natexlab{a}}, \aap, 595, A1

\bibitem[{{{\it Gaia} Collaboration} {et~al.}(2016{\natexlab{b}}){{\it Gaia}
  Collaboration}, {Brown}, {Vallenari}, {Prusti}, {de Bruijne}, {Mignard},
  {Drimmel}, {Babusiaux}, {Bailer-Jones}, {Bastian}, \& et~al.}]{gaia2016DR1}
{{\it Gaia} Collaboration}, {Brown}, A.~G.~A., {Vallenari}, A., {et~al.}
  2016{\natexlab{b}}, \aap, 595, A2

\bibitem[{{Jaffe} {et~al.}(2016){Jaffe}, {Barnes}, {Brooks}, {Lee}, {Mace},
  {Pak}, {Park}, \& {Park}}]{jaffe2016}
{Jaffe}, D.~T., {Barnes}, S., {Brooks}, C., {et~al.} 2016, in \procspie, Vol.
  9908, Ground-based and Airborne Instrumentation for Astronomy VI, 990821

\bibitem[{{Jeans}(1924)}]{jeans1924}
{Jeans}, J.~H. 1924, \mnras, 85, 2

\bibitem[{{Jenkins} {et~al.}(2009){Jenkins}, {Ramsey}, {Jones}, {Pavlenko},
  {Gallardo}, {Barnes}, \& {Pinfield}}]{jenkins2009}
{Jenkins}, J.~S., {Ramsey}, L.~W., {Jones}, H.~R.~A., {et~al.} 2009, \apj, 704,
  975

\bibitem[{{Johns-Krull} \& {Valenti}(1996)}]{johnskrull1996}
{Johns-Krull}, C.~M., \& {Valenti}, J.~A. 1996, \apjl, 459, L95

\bibitem[{{Johns-Krull} \& {Valenti}(2000)}]{johnskrull2000}
{Johns-Krull}, C.~M., \& {Valenti}, J.~A. 2000, in Astronomical Society of the
  Pacific Conference Series, Vol. 198, Stellar Clusters and Associations:
  Convection, Rotation, and Dynamos, ed. R.~{Pallavicini}, G.~{Micela}, \&
  S.~{Sciortino}, 371

\bibitem[{{Joy}(1947)}]{Joy1947}
{Joy}, A.~H. 1947, \apj, 105, 96

\bibitem[{{Kepler} {et~al.}(2016){Kepler}, {Pelisoli}, {Koester}, {Ourique},
  {Romero}, {Reindl}, {Kleinman}, {Eisenstein}, {Valois}, \&
  {Amaral}}]{kepler2016}
{Kepler}, S.~O., {Pelisoli}, I., {Koester}, D., {et~al.} 2016, \mnras, 455,
  3413

\bibitem[{{Kirkpatrick} {et~al.}(2012){Kirkpatrick}, {Gelino}, {Cushing},
  {Mace}, {Griffith}, {Skrutskie}, {Marsh}, {Wright}, {Eisenhardt}, {McLean},
  {Mainzer}, {Burgasser}, {Tinney}, {Parker}, \& {Salter}}]{kirkpatrick2012}
{Kirkpatrick}, J.~D., {Gelino}, C.~R., {Cushing}, M.~C., {et~al.} 2012, \apj,
  753, 156

\bibitem[{{Kochanek} {et~al.}(2017){Kochanek}, {Shappee}, {Stanek}, {Holoien},
  {Thompson}, {Prieto}, {Dong}, {Shields}, {Will}, {Britt}, {Perzanowski}, \&
  {Pojma{\'n}ski}}]{Kochanek2017}
{Kochanek}, C.~S., {Shappee}, B.~J., {Stanek}, K.~Z., {et~al.} 2017, \pasp,
  129, 104502

\bibitem[{{Kolb} \& {Stehle}(1996)}]{kolb1996}
{Kolb}, U., \& {Stehle}, R. 1996, \mnras, 282, 1454

\bibitem[{{Kouwenhoven} {et~al.}(2010){Kouwenhoven}, {Goodwin}, {Parker},
  {Davies}, {Malmberg}, \& {Kroupa}}]{Kouwenhoven2010}
{Kouwenhoven}, M.~B.~N., {Goodwin}, S.~P., {Parker}, R.~J., {et~al.} 2010,
  \mnras, 404, 1835

\bibitem[{{Kratter} \& {Matzner}(2006)}]{kratter2006}
{Kratter}, K.~M., \& {Matzner}, C.~D. 2006, \mnras, 373, 1563

\bibitem[{{Kundra} \& {Hric}(2011)}]{Kundra2011}
{Kundra}, E., \& {Hric}, L. 2011, \apss, 331, 121

\bibitem[{{Kurucz}(1979)}]{kurucz1979}
{Kurucz}, R.~L. 1979, \apjs, 40, 1

\bibitem[{{Kurucz}(2011)}]{kurucz11}
---. 2011, Canadian Journal of Physics, 89, 417

\bibitem[{{Lawrence} {et~al.}(2007){Lawrence}, {Warren}, {Almaini}, {Edge},
  {Hambly}, {Jameson}, {Lucas}, {Casali}, {Adamson}, {Dye}, {Emerson},
  {Foucaud}, {Hewett}, {Hirst}, {Hodgkin}, {Irwin}, {Lodieu}, {McMahon},
  {Simpson}, {Smail}, {Mortlock}, \& {Folger}}]{lawrence2007}
{Lawrence}, A., {Warren}, S.~J., {Almaini}, O., {et~al.} 2007, \mnras, 379,
  1599

\bibitem[{Lee \& Gullikson(2016)}]{jlee2016}
Lee, J.-J., \& Gullikson, K. 2016, plp: v2.1 alpha 3, , ,
  doi:10.5281/zenodo.56067.
\newblock \url{https://doi.org/10.5281/zenodo.56067}

\bibitem[{{Leggett}(1992)}]{leggett1992}
{Leggett}, S.~K. 1992, \apjs, 82, 351

\bibitem[{{L{\'e}pine} \& {Gaidos}(2011)}]{Lepine2011}
{L{\'e}pine}, S., \& {Gaidos}, E. 2011, \aj, 142, 138

\bibitem[{{L{\'e}pine} {et~al.}(2013){L{\'e}pine}, {Hilton}, {Mann}, {Wilde},
  {Rojas-Ayala}, {Cruz}, \& {Gaidos}}]{lepine2013}
{L{\'e}pine}, S., {Hilton}, E.~J., {Mann}, A.~W., {et~al.} 2013, \aj, 145, 102

\bibitem[{{L{\'e}pine} {et~al.}(2007){L{\'e}pine}, {Rich}, \&
  {Shara}}]{lepine2007}
{L{\'e}pine}, S., {Rich}, R.~M., \& {Shara}, M.~M. 2007, \apj, 669, 1235

\bibitem[{{Lindegren} {et~al.}(2016){Lindegren}, {Lammers}, {Bastian},
  {Hern{\'a}ndez}, {Klioner}, {Hobbs}, {Bombrun}, {Michalik}, {Ramos-Lerate},
  {Butkevich}, {Comoretto}, {Joliet}, {Holl}, {Hutton}, {Parsons},
  {Steidelm{\"u}ller}, {Abbas}, {Altmann}, {Andrei}, {Anton}, {Bach},
  {Barache}, {Becciani}, {Berthier}, {Bianchi}, {Biermann}, {Bouquillon},
  {Bourda}, {Br{\"u}semeister}, {Bucciarelli}, {Busonero}, {Carlucci},
  {Casta{\~n}eda}, {Charlot}, {Clotet}, {Crosta}, {Davidson}, {de Felice},
  {Drimmel}, {Fabricius}, {Fienga}, {Figueras}, {Fraile}, {Gai}, {Garralda},
  {Geyer}, {Gonz{\'a}lez-Vidal}, {Guerra}, {Hambly}, {Hauser}, {Jordan},
  {Lattanzi}, {Lenhardt}, {Liao}, {L{\"o}ffler}, {McMillan}, {Mignard}, {Mora},
  {Morbidelli}, {Portell}, {Riva}, {Sarasso}, {Serraller}, {Siddiqui}, {Smart},
  {Spagna}, {Stampa}, {Steele}, {Taris}, {Torra}, {van Reeven}, {Vecchiato},
  {Zschocke}, {de Bruijne}, {Gracia}, {Raison}, {Lister}, {Marchant},
  {Messineo}, {Soffel}, {Osorio}, {de Torres}, \& {O'Mullane}}]{lindgren2016}
{Lindegren}, L., {Lammers}, U., {Bastian}, U., {et~al.} 2016, \aap, 595, A4

\bibitem[{{Line} {et~al.}(2017){Line}, {Marley}, {Liu}, {Burningham}, {Morley},
  {Hinkel}, {Teske}, {Fortney}, {Freedman}, \& {Lupu}}]{line2017}
{Line}, M.~R., {Marley}, M.~S., {Liu}, M.~C., {et~al.} 2017, \apj, 848, 83

\bibitem[{{Mace} {et~al.}(2016{\natexlab{a}}){Mace}, {Jaffe}, {Park}, \&
  {Lee}}]{mace2016a}
{Mace}, G., {Jaffe}, D., {Park}, C., \& {Lee}, J.-J. 2016{\natexlab{a}}, in
  19th Cambridge Workshop on Cool Stars, Stellar Systems, and the Sun (CS19),
  55

\bibitem[{{Mace} {et~al.}(2016{\natexlab{b}}){Mace}, {Kim}, {Jaffe}, {Park},
  {Lee}, {Kaplan}, {Yu}, {Yuk}, {Chun}, {Pak}, {Kim}, {Lee}, {Sneden}, {Afsar},
  {Pavel}, {Lee}, {Oh}, {Jeong}, {Park}, {Kidder}, {Lee}, {Nguyen Le},
  {McLane}, {Gully-Santiago}, {Oh}, {Lee}, {Hwang}, \& {Park}}]{mace2016b}
{Mace}, G., {Kim}, H., {Jaffe}, D.~T., {et~al.} 2016{\natexlab{b}}, in
  \procspie, Vol. 9908, Society of Photo-Optical Instrumentation Engineers
  (SPIE) Conference Series, 99080C

\bibitem[{{Mace}(2014)}]{mace2014}
{Mace}, G.~N. 2014, PhD thesis, University of California, Los Angeles

\bibitem[{{Mace} {et~al.}(2013{\natexlab{a}}){Mace}, {Kirkpatrick}, {Cushing},
  {Gelino}, {Griffith}, {Skrutskie}, {Marsh}, {Wright}, {Eisenhardt}, {McLean},
  {Thompson}, {Mix}, {Bailey}, {Beichman}, {Bloom}, {Burgasser}, {Fortney},
  {Hinz}, {Knox}, {Lowrance}, {Marley}, {Morley}, {Rodigas}, {Saumon},
  {Sheppard}, \& {Stock}}]{mace2013a}
{Mace}, G.~N., {Kirkpatrick}, J.~D., {Cushing}, M.~C., {et~al.}
  2013{\natexlab{a}}, \apjs, 205, 6

\bibitem[{{Mace} {et~al.}(2013{\natexlab{b}}){Mace}, {Kirkpatrick}, {Cushing},
  {Gelino}, {McLean}, {Logsdon}, {Wright}, {Skrutskie}, {Beichman},
  {Eisenhardt}, \& {Kulas}}]{mace2013b}
---. 2013{\natexlab{b}}, \apj, 777, 36

\bibitem[{{Mahmud} {et~al.}(2011){Mahmud}, {Crockett}, {Johns-Krull}, {Prato},
  {Hartigan}, {Jaffe}, \& {Beichman}}]{mahmud2011}
{Mahmud}, N.~I., {Crockett}, C.~J., {Johns-Krull}, C.~M., {et~al.} 2011, \apj,
  736, 123

\bibitem[{{Maldonado} {et~al.}(2010){Maldonado}, {Mart{\'{\i}}nez-Arn{\'a}iz},
  {Eiroa}, {Montes}, \& {Montesinos}}]{maldonado2010}
{Maldonado}, J., {Mart{\'{\i}}nez-Arn{\'a}iz}, R.~M., {Eiroa}, C., {Montes},
  D., \& {Montesinos}, B. 2010, \aap, 521, A12

\bibitem[{{Mann} {et~al.}(2013{\natexlab{a}}){Mann}, {Brewer}, {Gaidos},
  {L{\'e}pine}, \& {Hilton}}]{mann2013b}
{Mann}, A.~W., {Brewer}, J.~M., {Gaidos}, E., {L{\'e}pine}, S., \& {Hilton},
  E.~J. 2013{\natexlab{a}}, \aj, 145, 52

\bibitem[{{Mann} {et~al.}(2015){Mann}, {Feiden}, {Gaidos}, {Boyajian}, \& {von
  Braun}}]{mann2015}
{Mann}, A.~W., {Feiden}, G.~A., {Gaidos}, E., {Boyajian}, T., \& {von Braun},
  K. 2015, \apj, 804, 64

\bibitem[{{Mann} {et~al.}(2013{\natexlab{b}}){Mann}, {Gaidos}, \&
  {Ansdell}}]{mann2013a}
{Mann}, A.~W., {Gaidos}, E., \& {Ansdell}, M. 2013{\natexlab{b}}, \apj, 779,
  188

\bibitem[{{Mann} {et~al.}(2016){Mann}, {Gaidos}, {Mace}, {Johnson}, {Bowler},
  {LaCourse}, {Jacobs}, {Vanderburg}, {Kraus}, {Kaplan}, \&
  {Jaffe}}]{mann2016a}
{Mann}, A.~W., {Gaidos}, E., {Mace}, G.~N., {et~al.} 2016, \apj, 818, 46

\bibitem[{{Maoz} {et~al.}(2010){Maoz}, {Sharon}, \& {Gal-Yam}}]{maoz2010}
{Maoz}, D., {Sharon}, K., \& {Gal-Yam}, A. 2010, \apj, 722, 1879

\bibitem[{{Marquardt} {et~al.}(2015){Marquardt}, {Sim}, {Ruiter}, {Seitenzahl},
  {Ohlmann}, {Kromer}, {Pakmor}, \& {R{\"o}pke}}]{marquardt2015}
{Marquardt}, K.~S., {Sim}, S.~A., {Ruiter}, A.~J., {et~al.} 2015, \aap, 580,
  A118

\bibitem[{{Marsh} {et~al.}(2007){Marsh}, {Mar}, \& {Jaffe}}]{marsh2007}
{Marsh}, J.~P., {Mar}, D.~J., \& {Jaffe}, D.~T. 2007, \ao, 46, 3400

\bibitem[{{Martin} {et~al.}(2005){Martin}, {Fanson}, {Schiminovich},
  {Morrissey}, {Friedman}, {Barlow}, {Conrow}, {Grange}, {Jelinsky},
  {Milliard}, {Siegmund}, {Bianchi}, {Byun}, {Donas}, {Forster}, {Heckman},
  {Lee}, {Madore}, {Malina}, {Neff}, {Rich}, {Small}, {Surber}, {Szalay},
  {Welsh}, \& {Wyder}}]{Martin2005}
{Martin}, D.~C., {Fanson}, J., {Schiminovich}, D., {et~al.} 2005, \apjl, 619,
  L1

\bibitem[{{Martin} {et~al.}(2017){Martin}, {Mace}, {McLean}, {Logsdon}, {Rice},
  {Kirkpatrick}, {Burgasser}, {McGovern}, \& {Prato}}]{martin2017}
{Martin}, E.~C., {Mace}, G.~N., {McLean}, I.~S., {et~al.} 2017, \apj, 838, 73

\bibitem[{{Maxted} {et~al.}(2004){Maxted}, {Marsh}, {Morales-Rueda}, {Barstow},
  {Dobbie}, {Schreiber}, {Dhillon}, \& {Brinkworth}}]{maxted2004}
{Maxted}, P.~F.~L., {Marsh}, T.~R., {Morales-Rueda}, L., {et~al.} 2004, \mnras,
  355, 1143

\bibitem[{{Meschiari} \& {Laughlin}(2010)}]{meschiari2010}
{Meschiari}, S., \& {Laughlin}, G.~P. 2010, \apj, 718, 543

\bibitem[{{Meschiari} {et~al.}(2009){Meschiari}, {Wolf}, {Rivera}, {Laughlin},
  {Vogt}, \& {Butler}}]{meschiari2009}
{Meschiari}, S., {Wolf}, A.~S., {Rivera}, E., {et~al.} 2009, \pasp, 121, 1016

\bibitem[{{Monteiro} {et~al.}(2006){Monteiro}, {Jao}, {Henry}, {Subasavage}, \&
  {Beaulieu}}]{monteiro2006}
{Monteiro}, H., {Jao}, W.-C., {Henry}, T., {Subasavage}, J., \& {Beaulieu}, T.
  2006, \apj, 638, 446

\bibitem[{{Muirhead} {et~al.}(2012){Muirhead}, {Hamren}, {Schlawin},
  {Rojas-Ayala}, {Covey}, \& {Lloyd}}]{muirhead2012}
{Muirhead}, P.~S., {Hamren}, K., {Schlawin}, E., {et~al.} 2012, \apjl, 750, L37

\bibitem[{{Muirhead} {et~al.}(2013){Muirhead}, {Vanderburg}, {Shporer},
  {Becker}, {Swift}, {Lloyd}, {Fuller}, {Zhao}, {Hinkley}, {Pineda}, {Bottom},
  {Howard}, {von Braun}, {Boyajian}, {Law}, {Baranec}, {Riddle}, {Ramaprakash},
  {Tendulkar}, {Bui}, {Burse}, {Chordia}, {Das}, {Dekany}, {Punnadi}, \&
  {Johnson}}]{muirhead2013}
{Muirhead}, P.~S., {Vanderburg}, A., {Shporer}, A., {et~al.} 2013, \apj, 767,
  111

\bibitem[{{Naud} {et~al.}(2014){Naud}, {Artigau}, {Malo}, {Albert}, {Doyon},
  {Lafreni{\`e}re}, {Gagn{\'e}}, {Saumon}, {Morley}, {Allard}, {Homeier},
  {Beichman}, {Gelino}, \& {Boucher}}]{naud2014}
{Naud}, M.-E., {Artigau}, {\'E}., {Malo}, L., {et~al.} 2014, \apj, 787, 5

\bibitem[{{Neves} {et~al.}(2013){Neves}, {Bonfils}, {Santos}, {Delfosse},
  {Forveille}, {Allard}, \& {Udry}}]{neves2013}
{Neves}, V., {Bonfils}, X., {Santos}, N.~C., {et~al.} 2013, \aap, 551, A36

\bibitem[{{Neves} {et~al.}(2012){Neves}, {Bonfils}, {Santos}, {Delfosse},
  {Forveille}, {Allard}, {Nat{\'a}rio}, {Fernandes}, \& {Udry}}]{neves2012}
---. 2012, \aap, 538, A25

\bibitem[{{Newton} {et~al.}(2014){Newton}, {Charbonneau}, {Irwin},
  {Berta-Thompson}, {Rojas-Ayala}, {Covey}, \& {Lloyd}}]{newton2014}
{Newton}, E.~R., {Charbonneau}, D., {Irwin}, J., {et~al.} 2014, \aj, 147, 20

\bibitem[{{Newton} {et~al.}(2017){Newton}, {Irwin}, {Charbonneau}, {Berlind},
  {Calkins}, \& {Mink}}]{newton2017}
{Newton}, E.~R., {Irwin}, J., {Charbonneau}, D., {et~al.} 2017, \apj, 834, 85

\bibitem[{{Newton} {et~al.}(2016){Newton}, {Irwin}, {Charbonneau},
  {Berta-Thompson}, {Dittmann}, \& {West}}]{newton2016}
---. 2016, \apj, 821, 93

\bibitem[{{Nidever} {et~al.}(2002){Nidever}, {Marcy}, {Butler}, {Fischer}, \&
  {Vogt}}]{nidever2002}
{Nidever}, D.~L., {Marcy}, G.~W., {Butler}, R.~P., {Fischer}, D.~A., \& {Vogt},
  S.~S. 2002, \apjs, 141, 503

\bibitem[{{Park} {et~al.}(2014){Park}, {Jaffe}, {Yuk}, {Chun}, {Pak}, {Kim},
  {Pavel}, {Lee}, {Oh}, {Jeong}, {Sim}, {Lee}, {Nguyen Le}, {Strubhar},
  {Gully-Santiago}, {Oh}, {Cha}, {Moon}, {Park}, {Brooks}, {Ko}, {Han}, {Nah},
  {Hill}, {Lee}, {Barnes}, {Yu}, {Kaplan}, {Mace}, {Kim}, {Lee}, {Hwang}, \&
  {Park}}]{park2014}
{Park}, C., {Jaffe}, D.~T., {Yuk}, I.-S., {et~al.} 2014, in \procspie, Vol.
  9147, Ground-based and Airborne Instrumentation for Astronomy V, 91471D

\bibitem[{{Parker} \& {Reggiani}(2013)}]{parker2013}
{Parker}, R.~J., \& {Reggiani}, M.~M. 2013, \mnras, 432, 2378

\bibitem[{{Plavchan} {et~al.}(2008){Plavchan}, {Jura}, {Kirkpatrick}, {Cutri},
  \& {Gallagher}}]{plavchan2008}
{Plavchan}, P., {Jura}, M., {Kirkpatrick}, J.~D., {Cutri}, R.~M., \&
  {Gallagher}, S.~C. 2008, \apjs, 175, 191

\bibitem[{{Prato} {et~al.}(2008){Prato}, {Huerta}, {Johns-Krull}, {Mahmud},
  {Jaffe}, \& {Hartigan}}]{prato2008}
{Prato}, L., {Huerta}, M., {Johns-Krull}, C.~M., {et~al.} 2008, \apjl, 687,
  L103

\bibitem[{{Prato} {et~al.}(2015){Prato}, {Mace}, {Rice}, {McLean},
  {Kirkpatrick}, {Burgasser}, \& {Kim}}]{prato2015}
{Prato}, L., {Mace}, G.~N., {Rice}, E.~L., {et~al.} 2015, \apj, 808, 12

\bibitem[{{Preibisch} {et~al.}(1999){Preibisch}, {Balega}, {Hofmann},
  {Weigelt}, \& {Zinnecker}}]{preibisch1999}
{Preibisch}, T., {Balega}, Y., {Hofmann}, K.-H., {Weigelt}, G., \& {Zinnecker},
  H. 1999, \na, 4, 531

\bibitem[{{Rayner} {et~al.}(2003){Rayner}, {Toomey}, {Onaka}, {Denault},
  {Stahlberger}, {Vacca}, {Cushing}, \& {Wang}}]{rayner2003}
{Rayner}, J.~T., {Toomey}, D.~W., {Onaka}, P.~M., {et~al.} 2003, \pasp, 115,
  362

\bibitem[{{Reddy} {et~al.}(2006){Reddy}, {Lambert}, \& {Allende
  Prieto}}]{reddy2006}
{Reddy}, B.~E., {Lambert}, D.~L., \& {Allende Prieto}, C. 2006, \mnras, 367,
  1329

\bibitem[{{Reid}(1996)}]{reid1996}
{Reid}, I.~N. 1996, \aj, 111, 2000

\bibitem[{{Reid} {et~al.}(1995){Reid}, {Hawley}, \& {Gizis}}]{reid1995}
{Reid}, I.~N., {Hawley}, S.~L., \& {Gizis}, J.~E. 1995, \aj, 110, 1838

\bibitem[{{Reipurth} \& {Mikkola}(2012)}]{reipurth2012}
{Reipurth}, B., \& {Mikkola}, S. 2012, \nat, 492, 221

\bibitem[{{Reipurth} \& {Mikkola}(2015)}]{reipurth2015}
---. 2015, \aj, 149, 145

\bibitem[{{Ricker} {et~al.}(2014){Ricker}, {Winn}, {Vanderspek}, {Latham},
  {Bakos}, {Bean}, {Berta-Thompson}, {Brown}, {Buchhave}, {Butler}, {Butler},
  {Chaplin}, {Charbonneau}, {Christensen-Dalsgaard}, {Clampin}, {Deming},
  {Doty}, {De Lee}, {Dressing}, {Dunham}, {Endl}, {Fressin}, {Ge}, {Henning},
  {Holman}, {Howard}, {Ida}, {Jenkins}, {Jernigan}, {Johnson}, {Kaltenegger},
  {Kawai}, {Kjeldsen}, {Laughlin}, {Levine}, {Lin}, {Lissauer}, {MacQueen},
  {Marcy}, {McCullough}, {Morton}, {Narita}, {Paegert}, {Palle}, {Pepe},
  {Pepper}, {Quirrenbach}, {Rinehart}, {Sasselov}, {Sato}, {Seager},
  {Sozzetti}, {Stassun}, {Sullivan}, {Szentgyorgyi}, {Torres}, {Udry}, \&
  {Villasenor}}]{ricker2014}
{Ricker}, G.~R., {Winn}, J.~N., {Vanderspek}, R., {et~al.} 2014, in \procspie,
  Vol. 9143, Space Telescopes and Instrumentation 2014: Optical, Infrared, and
  Millimeter Wave, 914320

\bibitem[{{Ritter} \& {Kolb}(1998)}]{Ritter1998}
{Ritter}, H., \& {Kolb}, U. 1998, \aaps, 129, 83

\bibitem[{Rojas-Ayala {et~al.}(2012)Rojas-Ayala, Covey, Muirhead, \&
  Lloyd}]{rojas2012}
Rojas-Ayala, B., Covey, K.~R., Muirhead, P.~S., \& Lloyd, J.~P. 2012, The
  Astrophysical Journal, 748, 93.
\newblock \url{http://stacks.iop.org/0004-637X/748/i=2/a=93}

\bibitem[{Rothman {et~al.}(2013)Rothman, Gordon, Babikov, Barbe, Benner,
  Bernath, Birk, Bizzocchi, Boudon, Brown, Campargue, Chance, Cohen, Coudert,
  Devi, Drouin, Fayt, Flaud, Gamache, Harrison, Hartmann, Hill, Hodges,
  Jacquemart, Jolly, Lamouroux, LeRoy, Li, Long, Lyulin, Mackie, Massie,
  Mikhailenko, M"{u}ller, Naumenko, Nikitin, Orphal, Perevalov, Perrin,
  Polovtseva, Richard, Smith, Starikova, Sung, Tashkun, Tennyson, Toon,
  Tyuterev, \& Wagner}]{HITRAN2012}
Rothman, L.~S., Gordon, I.~E., Babikov, Y., {et~al.} 2013, Journal of
  Quantitative Spectroscopy and Radiative Transfer, 130, 4

\bibitem[{{Ruiz-Lapuente} {et~al.}(1993){Ruiz-Lapuente}, {Jeffery}, {Challis},
  {Filippenko}, {Kirshner}, {Ho}, {Schmidt}, {S{\'a}nchez}, \&
  {Canal}}]{Ruiz1993}
{Ruiz-Lapuente}, P., {Jeffery}, D.~J., {Challis}, P.~M., {et~al.} 1993, \nat,
  365, 728

\bibitem[{{Scalzo} {et~al.}(2014){Scalzo}, {Ruiter}, \& {Sim}}]{scalzo2014}
{Scalzo}, R.~A., {Ruiter}, A.~J., \& {Sim}, S.~A. 2014, \mnras, 445, 2535

\bibitem[{{Schlaufman} \& {Laughlin}(2010)}]{schlaufman2010}
{Schlaufman}, K.~C., \& {Laughlin}, G. 2010, \aap, 519, A105

\bibitem[{{Schmidt} {et~al.}(2009){Schmidt}, {Wallerstein}, {Woolf}, \&
  {Bean}}]{schmidt2009}
{Schmidt}, S.~J., {Wallerstein}, G., {Woolf}, V.~M., \& {Bean}, J.~L. 2009,
  \pasp, 121, 1083

\bibitem[{{Scholz}(2010)}]{scholz2010}
{Scholz}, R.-D. 2010, \aap, 515, A92

\bibitem[{{Schreiber} \& {G{\"a}nsicke}(2003)}]{schreiber2003}
{Schreiber}, M.~R., \& {G{\"a}nsicke}, B.~T. 2003, \aap, 406, 305

\bibitem[{{Shappee} {et~al.}(2014){Shappee}, {Prieto}, {Grupe}, {Kochanek},
  {Stanek}, {De Rosa}, {Mathur}, {Zu}, {Peterson}, {Pogge}, {Komossa}, {Im},
  {Jencson}, {Holoien}, {Basu}, {Beacom}, {Szczygie{\l}}, {Brimacombe},
  {Adams}, {Campillay}, {Choi}, {Contreras}, {Dietrich}, {Dubberley},
  {Elphick}, {Foale}, {Giustini}, {Gonzalez}, {Hawkins}, {Howell}, {Hsiao},
  {Koss}, {Leighly}, {Morrell}, {Mudd}, {Mullins}, {Nugent}, {Parrent},
  {Phillips}, {Pojmanski}, {Rosing}, {Ross}, {Sand}, {Terndrup}, {Valenti},
  {Walker}, \& {Yoon}}]{Shappee2014}
{Shappee}, B.~J., {Prieto}, J.~L., {Grupe}, D., {et~al.} 2014, \apj, 788, 48

\bibitem[{{Shkolnik} {et~al.}(2012){Shkolnik}, {Anglada-Escud{\'e}}, {Liu},
  {Bowler}, {Weinberger}, {Boss}, {Reid}, \& {Tamura}}]{shkolnik2012}
{Shkolnik}, E.~L., {Anglada-Escud{\'e}}, G., {Liu}, M.~C., {et~al.} 2012, \apj,
  758, 56

\bibitem[{Shporer(2017)}]{shporer2017}
Shporer, A. 2017, Publications of the Astronomical Society of the Pacific, 129,
  072001.
\newblock \url{http://stacks.iop.org/1538-3873/129/i=977/a=072001}

\bibitem[{{Shulyak} {et~al.}(2014){Shulyak}, {Reiners}, {Seemann}, {Kochukhov},
  \& {Piskunov}}]{shulyak2014}
{Shulyak}, D., {Reiners}, A., {Seemann}, U., {Kochukhov}, O., \& {Piskunov}, N.
  2014, \aap, 563, A35

\bibitem[{{Sills} {et~al.}(2000){Sills}, {Pinsonneault}, \&
  {Terndrup}}]{sills2000}
{Sills}, A., {Pinsonneault}, M.~H., \& {Terndrup}, D.~M. 2000, \apj, 534, 335

\bibitem[{{Skrutskie} {et~al.}(2006){Skrutskie}, {Cutri}, {Stiening},
  {Weinberg}, {Schneider}, {Carpenter}, {Beichman}, {Capps}, {Chester},
  {Elias}, {Huchra}, {Liebert}, {Lonsdale}, {Monet}, {Price}, {Seitzer},
  {Jarrett}, {Kirkpatrick}, {Gizis}, {Howard}, {Evans}, {Fowler}, {Fullmer},
  {Hurt}, {Light}, {Kopan}, {Marsh}, {McCallon}, {Tam}, {Van Dyk}, \&
  {Wheelock}}]{skrutskie2006}
{Skrutskie}, M.~F., {Cutri}, R.~M., {Stiening}, R., {et~al.} 2006, \aj, 131,
  1163

\bibitem[{{Sneden}(1973)}]{sneden1973}
{Sneden}, C.~A. 1973, PhD thesis, THE UNIVERSITY OF TEXAS AT AUSTIN.

\bibitem[{{Sokal} {et~al.}(2018){Sokal}, {Deen}, {Mace}, {Lee}, {Oh}, {Kim},
  {Kidder}, \& {Jaffe}}]{Sokal2018}
{Sokal}, K.~R., {Deen}, C.~P., {Mace}, G.~N., {et~al.} 2018, ArXiv e-prints,
  arXiv:1712.04785

\bibitem[{{Stamatellos} \& {Whitworth}(2011)}]{stamatellos2011}
{Stamatellos}, D., \& {Whitworth}, A. 2011, in European Physical Journal Web of
  Conferences, Vol.~16, European Physical Journal Web of Conferences, 05001

\bibitem[{{Stassun} {et~al.}(2017){Stassun}, {Oelkers}, {Pepper}, {Paegert},
  {De Lee}, {Torres}, {Latham}, {Muirhead}, {Dressing}, {Rojas-Ayala}, {Mann},
  {Fleming}, {Levine}, {Silvotti}, {Plavchan}, \& {the TESS Target Selection
  Working Group}}]{stassun2017}
{Stassun}, K.~G., {Oelkers}, R.~J., {Pepper}, J., {et~al.} 2017, ArXiv
  e-prints, arXiv:1706.00495

\bibitem[{{Stauffer} \& {Hartmann}(1986{\natexlab{a}})}]{stauffer1986}
{Stauffer}, J.~B., \& {Hartmann}, L.~W. 1986{\natexlab{a}}, \pasp, 98, 1233

\bibitem[{{Stauffer} \& {Hartmann}(1986{\natexlab{b}})}]{stauffer1986b}
{Stauffer}, J.~R., \& {Hartmann}, L.~W. 1986{\natexlab{b}}, \apjs, 61, 531

\bibitem[{{Toonen} {et~al.}(2017){Toonen}, {Hollands}, {G{\"a}nsicke}, \&
  {Boekholt}}]{toonen2017}
{Toonen}, S., {Hollands}, M., {G{\"a}nsicke}, B.~T., \& {Boekholt}, T. 2017,
  \aap, 602, A16

\bibitem[{{Vacca} {et~al.}(2003){Vacca}, {Cushing}, \& {Rayner}}]{vacca2003}
{Vacca}, W.~D., {Cushing}, M.~C., \& {Rayner}, J.~T. 2003, \pasp, 115, 389

\bibitem[{{Vaccaro} {et~al.}(2015){Vaccaro}, {Wilson}, {Van Hamme}, \&
  {Terrell}}]{Vaccaro2015}
{Vaccaro}, T.~R., {Wilson}, R.~E., {Van Hamme}, W., \& {Terrell}, D. 2015,
  \apj, 810, 157

\bibitem[{{van Altena} {et~al.}(1995){van Altena}, {Lee}, \&
  {Hoffleit}}]{vanaltena1995}
{van Altena}, W.~F., {Lee}, J.~T., \& {Hoffleit}, E.~D. 1995, {The general
  catalogue of trigonometric [stellar] parallaxes}

\bibitem[{{van Leeuwen}(2007)}]{vanleeuwen2007}
{van Leeuwen}, F. 2007, \aap, 474, 653

\bibitem[{{Warner}(1995)}]{warner1995}
{Warner}, B. 1995, Cambridge Astrophysics Series, 28

\bibitem[{{Wesemael} {et~al.}(1993){Wesemael}, {Greenstein}, {Liebert},
  {Lamontagne}, {Fontaine}, {Bergeron}, \& {Glaspey}}]{wesemael1993}
{Wesemael}, F., {Greenstein}, J.~L., {Liebert}, J., {et~al.} 1993, \pasp, 105,
  761

\bibitem[{{Wheeler}(2012)}]{wheeler2012}
{Wheeler}, J.~C. 2012, \apj, 758, 123

\bibitem[{{Wood}(1992)}]{wood1992}
{Wood}, M.~A. 1992, \apj, 386, 539

\bibitem[{{Woolf} {et~al.}(2009){Woolf}, {L{\'e}pine}, \&
  {Wallerstein}}]{woolf2009}
{Woolf}, V.~M., {L{\'e}pine}, S., \& {Wallerstein}, G. 2009, \pasp, 121, 117

\bibitem[{{Woolf} \& {Wallerstein}(2006)}]{woolf2006}
{Woolf}, V.~M., \& {Wallerstein}, G. 2006, \pasp, 118, 218

\bibitem[{{Wright} {et~al.}(2010){Wright}, {Eisenhardt}, {Mainzer}, {Ressler},
  {Cutri}, {Jarrett}, {Kirkpatrick}, {Padgett}, {McMillan}, {Skrutskie},
  {Stanford}, {Cohen}, {Walker}, {Mather}, {Leisawitz}, {Gautier}, {McLean},
  {Benford}, {Lonsdale}, {Blain}, {Mendez}, {Irace}, {Duval}, {Liu}, {Royer},
  {Heinrichsen}, {Howard}, {Shannon}, {Kendall}, {Walsh}, {Larsen}, {Cardon},
  {Schick}, {Schwalm}, {Abid}, {Fabinsky}, {Naes}, \& {Tsai}}]{wright2010}
{Wright}, E.~L., {Eisenhardt}, P.~R.~M., {Mainzer}, A.~K., {et~al.} 2010, \aj,
  140, 1868

\bibitem[{{Wright} {et~al.}(2013){Wright}, {Skrutskie}, {Kirkpatrick},
  {Gelino}, {Griffith}, {Marsh}, {Jarrett}, {Nelson}, {Borish}, {Mace},
  {Mainzer}, {Eisenhardt}, {McLean}, {Tobin}, \& {Cushing}}]{wright2013}
{Wright}, E.~L., {Skrutskie}, M.~F., {Kirkpatrick}, J.~D., {et~al.} 2013, \aj,
  145, 84

\bibitem[{{Wu} \& {Wickramasinghe}(1993)}]{wu1993}
{Wu}, K., \& {Wickramasinghe}, D.~T. 1993, \mnras, 260, 141

\bibitem[{{Yuk} {et~al.}(2010){Yuk}, {Jaffe}, {Barnes}, {Chun}, {Park}, {Lee},
  {Lee}, {Wang}, {Park}, {Pak}, {Strubhar}, {Deen}, {Oh}, {Seo}, {Pyo}, {Park},
  {Lacy}, {Goertz}, {Rand}, \& {Gully-Santiago}}]{yuk2010}
{Yuk}, I.-S., {Jaffe}, D.~T., {Barnes}, S., {et~al.} 2010, in \procspie, Vol.
  7735, Ground-based and Airborne Instrumentation for Astronomy III, 77351M

\bibitem[{{Zahn}(1975)}]{zahn1975}
{Zahn}, J.-P. 1975, \aap, 41, 329

\bibitem[{{Zahn}(1977)}]{zahn1977}
---. 1977, \aap, 57, 383

\bibitem[{{Zahn} \& {Bouchet}(1989)}]{zahn1989}
{Zahn}, J.-P., \& {Bouchet}, L. 1989, \aap, 223, 112

\bibitem[{{Zinnecker} \& {Yorke}(2007)}]{zinnecker2007}
{Zinnecker}, H., \& {Yorke}, H.~W. 2007, \araa, 45, 481

\bibitem[{{Zucker} {et~al.}(2007){Zucker}, {Mazeh}, \&
  {Alexander}}]{Zucker2007}
{Zucker}, S., {Mazeh}, T., \& {Alexander}, T. 2007, \apj, 670, 1326

\bibitem[{{Zuckerman} \& {Becklin}(1987)}]{zuckerman1987}
{Zuckerman}, B., \& {Becklin}, E.~E. 1987, \apjl, 319, L99

\end{thebibliography}

\begin{deluxetable}{lccccccl}	
\tabletypesize{\tiny} 
\tablewidth{0pt}							
\tablecaption{{\it V}$-$band Photometry of Wolf 1130AB							
\label{tbl-1}}							
\tablehead{							
\colhead{MJD} 	&	\colhead{{\it V}$-$band} 	&	 \colhead{$\sigma_{{\it V}}$}  	&	\colhead{Relative Phase\tablenotemark{a}} 	\\
 \colhead{} 	&	 \colhead{(mag.)} 	&	\colhead{(mag.)} 	&               }	
\startdata													
56857.16630	&	11.849	&	0.003	&	0.951	\\
56857.16745	&	11.851	&	0.003	&	0.953	\\
56857.16860	&	11.851	&	0.003	&	0.955	\\
56857.16976	&	11.852	&	0.003	&	0.958	\\
56857.17091	&	11.850	&	0.003	&	0.960	\\
56857.17206	&	11.850	&	0.003	&	0.962	\\
56857.17322	&	11.849	&	0.003	&	0.965	\\
56857.17438	&	11.854	&	0.003	&	0.967	\\
56857.17553	&	11.849	&	0.003	&	0.969	\\
56857.17668	&	11.851	&	0.003	&	0.971	\\
56857.17783	&	11.852	&	0.003	&	0.974	\\
56857.17899	&	11.849	&	0.003	&	0.976	\\
56857.18813	&	11.852	&	0.003	&	0.995	\\
56857.18928	&	11.852	&	0.003	&	0.997	\\
56857.19044	&	11.856	&	0.003	&	0.999	\\
56857.19159	&	11.856	&	0.003	&	0.002	\\
56857.19274	&	11.858	&	0.003	&	0.004	\\
56857.19714	&	11.858	&	0.003	&	0.013	\\
56857.19829	&	11.854	&	0.003	&	0.015	\\
56857.19945	&	11.852	&	0.003	&	0.017	\\
\enddata		

\tablenotetext{a}{In this work phase$=$0 at inferior conjunction.}

\end{deluxetable}

\begin{deluxetable}{lccccccl}	
\tabletypesize{\tiny} 
\tablewidth{0pt}									
\tablecaption{IGRINS Derived Radial Velocities of Wolf 1130A									
\label{tbl-2}}									
\tablehead{									
\colhead{UT Date} 	&	\colhead{MJD} 	&	\colhead{$v_1$}  	&	 \colhead{$\sigma_{v1}$}  	&	\colhead{Relative Phase\tablenotemark{a}} 	\\
\colhead{YYYYMMDD} 	&	 \colhead{} 	&	\colhead{(km s$^{-1}$)} 	&	 \colhead{(km s$^{-1}$)} 	&	}		
\startdata									
20140711	&	56849.3578	&	84.08	&	0.18	&	0.230	\\
20140924	&	56924.0988	&	-151.07	&	0.16	&	0.704	\\
20141010	&	56940.0670	&	-130.21	&	0.16	&	0.852	\\
20141010	&	56940.1324	&	-44.61	&	0.16	&	0.984	\\
20141010	&	56940.2080	&	58.04	&	0.16	&	0.136	\\
20141011	&	56941.0636	&	-127.31	&	0.16	&	0.859	\\
20141018	&	56948.2753	&	49.26	&	0.17	&	0.378	\\
20141125	&	56986.1621	&	-132.98	&	0.16	&	0.654	\\
20141125	&	56986.1758	&	-144.01	&	0.16	&	0.682	\\
20141125	&	56986.1908	&	-150.39	&	0.16	&	0.712	\\
20141126	&	56987.0405	&	25.71	&	0.16	&	0.423	\\
20141126	&	56987.0554	&	2.68	&	0.16	&	0.452	\\
20141126	&	56987.0864	&	-43.64	&	0.16	&	0.515	\\
20141126	&	56987.1020	&	-66.29	&	0.16	&	0.546	\\
20141126	&	56987.1123	&	-87.21	&	0.16	&	0.567	\\
20141126	&	56987.1471	&	-129.06	&	0.16	&	0.637	\\
20141126	&	56987.1643	&	-140.73	&	0.16	&	0.672	\\
20141126	&	56987.1781	&	-148.86	&	0.16	&	0.700	\\
20150611	&	57184.3367	&	-123.18	&	0.16	&	0.633	\\
20150611	&	57184.3931	&	-154.80	&	0.16	&	0.747	\\
20150611	&	57184.4322	&	-141.43	&	0.16	&	0.825	\\
20150612	&	57185.3596	&	-146.87	&	0.16	&	0.693	\\
20150701	&	57204.3627	&	-70.09	&	0.16	&	0.951	\\
20150701	&	57204.3847	&	-38.69	&	0.16	&	0.995	\\
20150702	&	57205.3419	&	-89.56	&	0.16	&	0.922	\\
20150702	&	57205.3614	&	-61.59	&	0.16	&	0.962	\\
20150703	&	57206.2948	&	-135.68	&	0.16	&	0.841	\\
20150703	&	57206.3429	&	-79.37	&	0.16	&	0.938	\\
20150703	&	57206.4369	&	54.24	&	0.16	&	0.127	\\
20150703	&	57206.4610	&	74.78	&	0.16	&	0.175	\\
20150804	&	57238.2906	&	87.69	&	0.16	&	0.257	\\
20150804	&	57238.3074	&	82.75	&	0.16	&	0.291	\\
20150805	&	57239.2435	&	76.18	&	0.16	&	0.175	\\
20150805	&	57239.2697	&	87.09	&	0.16	&	0.228	\\
20150805	&	57239.2834	&	87.46	&	0.16	&	0.256	\\
20150805	&	57239.2974	&	84.98	&	0.16	&	0.284	\\
20150805	&	57239.3299	&	64.13	&	0.16	&	0.349	\\
20150805	&	57239.3445	&	49.53	&	0.16	&	0.379	\\
20150805	&	57239.3584	&	32.43	&	0.16	&	0.407	\\
20150805	&	57239.3894	&	-10.38	&	0.17	&	0.469	\\
\enddata				

\tablenotetext{a}{In this work phase$=$0 at inferior conjunction.}

\end{deluxetable}

\begin{rotatetable} 
\begin{deluxetable}{lcccccc}	
\tabletypesize{\tiny} 
\tablewidth{0pt}																					
\tablecaption{Orbital Elements and Derived Properties}																					
\label{tbl-3}																				
\tablehead{\colhead{$~$} 	&	\multicolumn{2}{c}{Individual Fit} 			&	\multicolumn{3}{c}{Combined Fit}\\													
\colhead{Element/Property} 	&	 \colhead{Visible Light}  			&	 \colhead{Infrared} 			&	 \colhead{Free-parameter} 			&	 \colhead{fixed \textit{e} = 0}  			&	 \colhead{fixed \textit{e} = 0, $i$ = 29}}
\startdata																	
P (days)	&	0.4967013	$\pm$	0.0000006	&	0.4967040	$\pm$	0.0000007	&	0.49670419	$\pm$	0.00000004	&	0.49670418	$\pm$	0.00000005	&	0.49670418	$\pm$	0.00000005	\\
$\gamma$ (\kms\ ) 	&	-34.1			&	-33.5			&	-33.2			&	-33.2			&	-33.2			\\
\textit{e} 	&	0.011	$\pm$	0.003	&	0.002	$\pm$	0.002	&	0.002	$\pm$	0.002	&	0\tablenotemark{a}			&	0\tablenotemark{a}			\\
$\omega$ (degrees) 	&	210	$\pm$	19	&	160	$\pm$	73	&	164	$\pm$	62	&	163	$\pm$	24	&	164	$\pm$	18	\\
T (MJD)	&	49559.04	$\pm$	0.03	&	56849.09	$\pm$	0.10	&	56849.10	$\pm$	0.09	&	56849.10	$\pm$	0.03	&	56848.942	$\pm$	0.03	\\
$M_A$ (M$_{\odot}$)	&	0.30\tablenotemark{a}			&	0.30\tablenotemark{a}			&	0.30\tablenotemark{a}			&	0.30\tablenotemark{a}			&	0.30\tablenotemark{a}			\\
$M_B$~sin~$i$ (M$_{\odot}$)	&	0.341	$\pm$	0.002	&	0.332	$\pm$	0.001	&	0.332	$\pm$	0.001	&	0.332	$\pm$	0.001	&	1.242	$\pm$	0.005\tablenotemark{b}	\\
a~sin~$i$ (au) 	&	0.01058	$\pm$	0.00001	&	0.010537	$\pm$	0.000005	&	0.010537	$\pm$	0.000005	&	0.010537	$\pm$	0.000005	&	0.014183	$\pm$	0.000015\tablenotemark{c}	\\
K (\kms\ )	&	123.3	$\pm$	0.5	&	121.3	$\pm$	0.2	&	121.3	$\pm$	0.2	&	121.3	$\pm$	0.2	&	121.3	$\pm$	0.2	\\
$\chi$~$_{reduced}^2$ 	&	0.38			&	70.52			&	40.61			&	41.54			&	42.25			\\
N &	27			&	40			&	67			&	67			&	67  \\
\enddata 
\tablenotetext{a}{This is a fixed parameter.}
\tablenotetext{b}{$i=$29 and this is $M_B$, the mass of the white dwarf.}
\tablenotetext{c}{$i=$29 and this is a, the orbital separation.}

\end{deluxetable} 
\clearpage 
\end{rotatetable}

\begin{deluxetable}{lcccc}		
\tabletypesize{\tiny} 
\tablewidth{0pt}			
\tablecaption{Physical Parameters for Wolf~1130A			
\label{tbl-4}}			
\tablehead{\colhead{$~$} 			&\colhead{$~$}\\
\colhead{Element/Property}			&\colhead{Measurement}}
\startdata			
{\it Gaia} Parallax (mas)	&	59.91$\pm$0.55	\\
Distance (pc)	&	16.69$\pm$0.15	\\
Parallax Radius\tablenotemark{a} (R$_{\odot}$) &	0.302$\pm$0.009	\\
Model Radius\tablenotemark{b} (R$_{\odot}$) & 0.289$\pm$0.011	\\
Parallax Mass\tablenotemark{c} (M$_{\odot}$)	& 0.308$\pm$0.016	\\
Model Mass\tablenotemark{b} (M$_{\odot}$) & 0.297$\pm$0.011	\\
$i$	($^{\circ}$) &	29$\pm$2	\\
v~sin~$i$ (\kms\ )	&	14.7$\pm$0.7  \\
T$_{\rm eff}$\tablenotemark{d} (K)	&	3530$\pm$60	\\
log~$g$\tablenotemark{d} (dex)	&	4.9	\\
B-field (kG)    &   $<$3  \\
Age (Gyr) & $>$10 \\
{[M/H]} (dex) &	$-$0.45$\pm$0.12\tablenotemark{e} \\
{[Fe/H]} (dex) & $-$0.70$\pm$0.12, $-$0.64$\pm$0.17\tablenotemark{e}, $-$0.62$\pm$0.10\tablenotemark{f}\\
{[Ti/H]} (dex) &	$-$0.22$\pm$0.09\tablenotemark{f}\\
{[O/H]} (dex) &	$-$0.5$\pm$0.1, $-$0.45$\pm$0.11\tablenotemark{g}\\
{[Ca/H]} (dex) &	$-$0.20$\pm$0.05\\
\enddata 

\tablenotetext{a}{M$_K$-Radius-[Fe/H] relationship from \citet{mann2015}.}
\tablenotetext{b}{Dartmouth Stellar Evolution Models \citep{dotter2008}.}
\tablenotetext{c}{M$_K$-Mass relationship from \citet{benedict2016}.}
\tablenotetext{d}{Weighted average determined from fit to MDM and STIS spectra.}
\tablenotetext{e}{\citet{rojas2012}}
\tablenotetext{f}{\citet{woolf2006}}
\tablenotetext{g}{\citet{schmidt2009}}

\end{deluxetable} 

\begin{rotatetable}
\center 
\begin{deluxetable}{lccccccl}	
\tabletypesize{\tiny} 
\tablewidth{0pt}									
\tablecaption{T Dwarf Multiple System Characteristics								
\label{tbl-5}}									
\tablehead{									
\colhead{Object Name} 	&	\colhead{Discovery Reference} 	&	\colhead{SpT}  	&	 \colhead{Age (Gyr)}  	&	\colhead{Separation (AU)} 	& \colhead{System Components}   & \colhead{Distance (pc)}}								
\startdata													
Gl~570D	&	\citet{burgasser2000}	&	T7.5	&   2$-$10	&	1450\tablenotemark{a}	&	K4+(M1.5+M3)	&	5.8\tablenotemark{b}	\\
Ross~458C 	&	\citet{goldman2010, burgasser2010, scholz2010}	&	T8.5	&	$<$1	&	1100	&	M0.5+M7	&	11.7\tablenotemark{b}	\\
$\xi$~UMa~E	&	\citet{wright2013}	&	T8.5 	&   2$-$8	&	4100	&	F8.5+G2\tablenotemark{c}	&	8.8\tablenotemark{b}	\\
Wolf~1130C	&	\citet{mace2013b}	&	T8	&	$>$10\tablenotemark{d}	&	3150	&	M3+WD	&	16.7	\\
2MASS~J0213+3648~C	&	\citet{deacon2017}	&	T3	&	1$-$10	&	360	&	M4.5+M6.5	&	22	\\
\enddata												

\tablenotetext{a}{Separation for Gl~570D is the average of the A-D and BC-D separations reported by \citet{burgasser2000}.}										
\tablenotetext{b}{Distance from the updated 20~pc sample in Table 8 of \citet{kirkpatrick2012}.}													
\tablenotetext{c}{Both listed components of $\xi$ Uma are spectroscopic binaries.}													
\tablenotetext{d}{Age based on M dwarf metallicity and kinematics. The non-detection of the white dwarf in this system produces an age bound $>$3.4~Gyrs.}				
\end{deluxetable} 
\clearpage 
\end{rotatetable}

\end{document}